\documentclass{article}
\usepackage{graphicx} 
\usepackage{cite}
\usepackage{amsmath}
\usepackage{physics}
\usepackage{notes2bib}
\usepackage{tikz-feynman}
\usepackage{subcaption}
\tikzfeynmanset{
every blob={/tikz/fill=white!30,/tikz/inner sep=2pt},
}
\numberwithin{equation}{section}

\title{Gluon Saturation\footnote{Submitted as a contribution to the ``Encyclopedia of Nuclear Physics''}}
\author{Alfred H. Mueller \\
Department of Physics\\
Columbia University\\
New York, NY 10027}
\date{June 2026}

\begin{document}

\maketitle

\section*{Abstract}

This is a review on gluon saturation. Sec.~\ref{sec:1} gives a general introduction to the subject while Sec.~\ref{sec:2} has a brief discussion of quark saturation. Sec.~\ref{sec:3} covers gluon saturation in a semiclassical limit and the much more general context of QCD evolution. Surprisingly gluon occupancies in a high energy hadron can become arbitrarily large, much larger than $1/\alpha$. Sudakov effects are introduced in Sec.~\ref{sec:4} and reduce the maximum gluon occupancies to less than or equal to $(1/\alpha)^{3/2}$ in both fixed coupling and running coupling QCD. In Sec.~\ref{sec:5} the role of saturation in coherent states in QCD is reviewed as well as brief discussions on nuclear shadowing, heavy ion collisions and Drell-Yan like processes are given.


\section{Introduction} \label{sec:1}

This is a discussion of gluon saturation and related topics \cite{cite1,cite2,cite3,cite4,cite5,cite6,cite7,cite8,cite9,cite10,cite11,cite12,cite13,cite14,cite15,cite16}. The focus is on understanding and appreciating saturation with most technical parts of the discussion not given completely, but with references given for the reader who wants more detail. Since quark and gluon saturation are related in many ways, the first chapter of this article, Sec.~\ref{sec:2}, is a brief discussion of quark saturation. Most of what follows will be focused on electron-nucleus scattering both inelastic, where the nucleus breaks up and coherent, or elastic, where the nucleus remains in its ground state after the scattering. In our discussion we shall focus on coherent scattering \cite{cite11,cite12} and coherent occupancies, but generally the corresponding inelastic
occupancies and TMD's follow rather easily from the coherent properties.

Quark saturation occurs when the quark-antiquark pair, coming from the $\gamma^*$ in electron-nucleus scattering, reaches the unitarity limit when scattering on, say, a large nucleus \cite{cite11,cite13}. In the scattering process which we call the projectile picture the incoming quark and antiquark have transverse momentum $p_\perp$ and $-p_\perp$ respectively and $p_\perp$ is one of the variables fixed in the quark TMD. The other momentum variable being fixed in the TMD is $p_-$ the fraction of the minus component of momentum (of a nucleon in the nucleus) carried by the quark before it is struck by the $\gamma^*$. (The $\gamma^*$ has momentum $q=(q_+,-\frac{Q^2}{q_+},\underline{0})$ while the nucleons in the nucleus have momentum $P = (\frac{m^2}{2P_-},P_-,0)$.) The TMD is then a function of $\frac{p_-}{P_-}$ and $p_\perp$ and when we refer to occupancies it will be in terms of these
variables. The scattering picture where the time variable is $x_+$ is illustrated in Fig.~\ref{fig:1} and Fig.~\ref{fig:2A} while the target picture in Fig.~\ref{fig:2B}. It is in the target picture with $x_-$ the time variable where the partonic picture is manifest in $A_-=0$ gauge.

For quarks in high energy QCD the maximum quark occupancy is given in \eqref{eq:2.13} and is parametrically on the order of one. This maximum quark occupancy occurs when the diquark-nuclear scattering in Fig.~\ref{fig:2A} reaches its unitary limit \cite{cite9,cite13}. At this point saturation implies strong leading twist shadowing \cite{cite10,cite17}.

Section \ref{sec:3} deals with the basics of gluon saturation. Just as quark saturation and unitarity are related to quark dipole scattering on a nucleus one can relate gluon saturation, and unitarity, to gluon dipole-nucleus scattering as illustrated 
in Fig.~\ref{fig:3} and Fig.~\ref{fig:10}. This is done as seen in Fig.~\ref{fig:10} by having the $\gamma^*$ break into a high transverse momentum $q\bar{q}$ pair which itself would interact weakly with a nucleus. We then require the $q\bar{q}$ pair to emit a reasonably soft gluon $k_\perp$. Then the $q\bar{q}g$ system scatters on the nucleus in the same way as a gluon dipole of momentum $k_\perp$ \cite{cite11,cite12}. Fig.~\ref{fig:10} shows the projectile picture while Fig.~\ref{fig:7} the target or partonic picture. The quark loop of Fig.~\ref{fig:7} is the measurer of the gluon TMD labeled by $k_\perp$ and $k_-$ with $k_-$ fixed by putting the $q\bar{q}$ pair on mass shell. As in the quark case the (coherent) gluon TMD is a function of $k_\perp$ and $k_-$. So we expect gluon saturation to be reached when the gluon dipole of Fig.~\ref{fig:10} reaches its unitarity limit when scattering off the nucleus exactly as was the case in quark saturation. Now however there
is a subtlety.

In the quark case fixing the $p_-$ and $p_\perp$ in Fig.~\ref{fig:2} effectively also fixes $p_+$. Now, however, fixing $k_-$ and $k_\perp$ in Fig.~\ref{fig:10} does not fix $k_+$. In fact there is a $\frac{\dd{k_+}}{k_+}$ logarithmic integral leading to \eqref{eq:3.15} rather than \eqref{eq:2.14}. The $N_c/4\pi^4$ factor for quark saturation (unitarity) becomes $(N_c^2-1)/8\pi^4$ in the gluon case as $N_c$ and $N_c^2-1$ count the number of colors in the two cases and the quark factor has an extra 2 because there are both quarks and antiquarks in the TMD of \eqref{eq:2.14}. However, the $(Y-\bar{Y})$ in \eqref{eq:3.15} is directly from the $\frac{\dd{k_+}}{k_+}$ integration because in the gluon case $k_+$ is not determined by knowing $k_-$ and $k_\perp$. It thus appears that gluon occupations can become very large because unitarity only fixes the gluon dipole-nucleus scattering in Fig.~\ref{fig:10} once 
$k_+$ has also been fixed.

The factor of $(Y-\bar{Y})$ in \eqref{eq:3.15} can be given in terms of the saturation momentum and one finally reaches the form given in \eqref{eq:3.17} where the gluon occupancy is proportional to $\frac{1}{\alpha}\ln\frac{Q_s^2}{k_\perp^2}$. The surprise here is that this is exactly the same form of the gluon occupancy as for quasiclassical gluons given in \eqref{eq:3.5}.

There is a mechanism for, partially, limiting gluon occupancies and that is through Sudakov effects \cite{cite18,cite19,cite20,cite21,cite22}. In the $\frac{\dd{k_+}}{k_+}$ integral which gives the large factor $(Y-\bar{Y})$ there is a large rapidity gap between the $q\bar{q}$ and the soft gluon $k$ of Fig.~\ref{fig:10}. We have required that this gap exist in our discussion, that is we have required that there be no gluon emissions at rapidities between that of gluon $k$ and that ``gluon'' $q-k$. This is unlikely and that
small probability is given by the Sudakov factor \eqref{eq:4.3}. In Sec.~\ref{sec:4.2} we put in running coupling effects but the final answer, including or not running coupling in the Sudakov factor and in the QCD evolution, does not distinguish between a fixed coupling calculation and a running coupling calculation with the gluon TMD given by \eqref{eq:4.20}. Now the occupancies increase as $k_\perp$ goes below $Q_s$ until an occupancy of $1/\alpha^{3/2}$ is reached which is the maxim occupancy for gluons in QCD.

In Sec.~\ref{sec:5.1} we note that the part of the nuclear wavefunction which leads to coherent reactions is, in fact, a coherent state for gluon pairs. This is made clear by looking at the projectile picture of scattering but 
is far from evident when looking directly at the target (partonic) frame. I think it is one of the interesting challenges to understand, directly in the target frame, how coherent pairs with zero color charge and zero total transverse momentum come about.

In Sec.~\ref{sec:5.2} we go into some detail on the relationship between saturation and shadowing \cite{cite17}. In the quark TMD quark saturation and shadowing are identical phenomena, both directly related to the unitarity limits in quark dipole-nucleus scattering. In the gluon case the relationship is more subtle. For the quasiclassical gluon there is no shadowing but TMD formulas \textit{look} like they are coming from unitarity because the formulas for $p_\perp$-broadening are identical to those for gluon dipole-nucleon scattering. Beyond the semiclassical limit the gluon TMD
is limited by unitarity effects in an identical way for shadowing and for occupancy and this is again a leading twist shadowing.

In Sec.~\ref{sec:5.3} we briefly discuss the saturated nucleus as determining the initial state in heavy ion collisions \cite{cite24,cite25,cite26}. We note that the important role is played by gluons having $k_\perp\sim Q_s$ and the very high occupancy gluons having $k_\perp/Q_s \ll 1$ play essentially no role in heavy ion collisions.

In Sec.~\ref{sec:5.4} we discuss the issue of gluon recombination as a mechanism for limiting the occupancy of gluons. In the end we find no evidence for recombination.

Finally in Sec.~\ref{sec:5.5} we remind the reader that the usual relationship between Drell-Yan $\mu$-pair production and the quark TMD from
DIS is changed by a factor of 4 at very high energies \cite{cite13} when $q_\perp$, the transverse momentum of the $\mu$-pairs, is less than $Q_s$. This is because the inelastic and coherent TMD's both contribute even though the nuclei always break up in the collision.

\section{Quark saturation} \label{sec:2}

Although the focus of this paper is on gluon saturation it is useful to begin our discussion with the simpler case of quark saturation where the discussion is more straightforward and where many of the subtleties of gluon saturation are absent. As we shall see saturation is intimately related to unitarity limits in corresponding scatting processes. And, indeed, we begin our discussion with the scattering of a $q\bar{q}$ dipole, coming from a $\gamma^*$, on a simple unevolved nucleus, a McLerran-Venugopalan nucleus \cite{cite9}.

\subsection{Projectile picture of coherent scattering} \label{sec:2.1}

The picture of the process is shown in Fig.~\ref{fig:1} where the arrows on the quark lines indicate the
direction of baryon number flow. The value of $Q^2=-q^2$ is taken very large but we suppose the transverse coordinate of the antiquark, $\underline{x}$ or $\underline{x}'$, is large enough so that the dipole, $(\underline{x},\underline{0})$ or $(\underline{x}',\underline{0})$, cross section on the nucleus is large. This is the standard aligned jet configuration. Then the scattering of the $(q\bar{q})$ dipole in the amplitude with a single nucleon of the nucleus is
\begin{equation} \label{eq:2.1}
    T(x_\perp) = \frac{\pi^2\alpha C_F}{N_c^2-1} xG(x,x_\perp^2) x_\perp^2
\end{equation}
where $xG$ corresponds to the nucleon gluon density at a scale $1/x_\perp^2$.

\begin{figure}[htbp]
\centering
\begin{tikzpicture}
\begin{feynman} 
\vertex (p1) {\(\gamma^*\)};
\vertex [right=1.5cm of p1] (l);
\vertex [below right=0.707cm of l] (b1);
\vertex [right=0.25cm of b1] (b2);
\vertex [right=1.5cm of b2] (b3);
\vertex [right=0.75cm of b3] (b4);
\vertex [right=0.75cm of b4] (b5);
\vertex [right=1.5cm of b5] (b6);
\vertex [right=0.25cm of b6] (b7);
\vertex [above right=0.707cm of b7] (r);
\vertex [right=1.5cm of r] (p2) {\(\gamma^*\)};
\vertex [above right=0.707cm of l] (t1);
\vertex [right=0.25cm of t1] (t2);
\vertex [right=1.5cm of t2] (t3);
\vertex [right=0.75cm of t3] (t4);
\vertex [right=0.75cm of t4] (t5);
\vertex [right=1.5cm of t5] (t6);
\vertex [right=0.25cm of t6] (t7);

\vertex [above=0.5cm of t4] (tt);
\vertex [below=0.5cm of b4] (bb);

\vertex [above=0.3cm of b2] (nt1);
\vertex [below=0.5cm of b2, blob, minimum height=0.5cm, minimum width=0.5cm] (n1) {};
\vertex [above=0.22cm of n1] (nb1);
\vertex [left=0.1cm of nt1] (ntl1);
\vertex [right=0.1cm of nt1] (ntr1);
\vertex [left=0.1cm of nb1] (nbl1);
\vertex [right=0.1cm of nb1] (nbr1);
\vertex [below left=0.5cm of n1] (nl1);
\vertex [below right=0.5cm of n1] (nr1);

\vertex [above=0.3cm of b3] (nt2);
\vertex [below=0.5cm of b3, blob, minimum height=0.5cm, minimum width=0.5cm] (n2) {};
\vertex [above=0.22cm of n2] (nb2);
\vertex [left=0.1cm of nt2] (ntl2);
\vertex [right=0.1cm of nt2] (ntr2);
\vertex [left=0.1cm of nb2] (nbl2);
\vertex [right=0.1cm of nb2] (nbr2);
\vertex [below left=0.5cm of n2] (nl2);
\vertex [below right=0.5cm of n2] (nr2);

\vertex [above=0.3cm of b5] (nt3);
\vertex [below=0.5cm of b5, blob, minimum height=0.5cm, minimum width=0.5cm] (n3) {};
\vertex [above=0.22cm of n3] (nb3);
\vertex [left=0.1cm of nt3] (ntl3);
\vertex [right=0.1cm of nt3] (ntr3);
\vertex [left=0.1cm of nb3] (nbl3);
\vertex [right=0.1cm of nb3] (nbr3);
\vertex [below left=0.5cm of n3] (nl3);
\vertex [below right=0.5cm of n3] (nr3);

\vertex [above=0.3cm of b6] (nt4);
\vertex [below=0.5cm of b6, blob, minimum height=0.5cm, minimum width=0.5cm] (n4) {};
\vertex [above=0.22cm of n4] (nb4);
\vertex [left=0.1cm of nt4] (ntl4);
\vertex [right=0.1cm of nt4] (ntr4);
\vertex [left=0.1cm of nb4] (nbl4);
\vertex [right=0.1cm of nb4] (nbr4);
\vertex [below left=0.5cm of n4] (nl4);
\vertex [below right=0.5cm of n4] (nr4);

\diagram* {
  (l) -- [quarter right] (b1) -- (b2) -- [edge label=\(z\)] (b3) -- (b4) -- [fermion, edge label=\(\underline{0}\)] (b5) -- (b6) -- (b7) -- [quarter right] (r),
  (l) -- [quarter left] (t1) -- (t2) -- [edge label=\(1-z\)] (t3) -- [anti fermion, edge label=\(\underline{x}\)] (t4) -- [edge label=\(\underline{x}'\)] (t5) -- (t7) -- [quarter left] (r),
  (p1) -- [charged boson, edge label'=\(q\)] (l),
  (r) -- [charged boson, edge label'=\(q\)] (p2),
  
  (ntl1) -- [photon, double] (nbl1),
  (ntr1) -- [photon, double] (nbr1),
  (nl1) -- [double] (n1) -- [double] (nr1),
  
  (ntl2) -- [photon, double] (nbl2),
  (ntr2) -- [photon, double] (nbr2),
  (nl2) -- [double] (n2) -- [double] (nr2),
  
  (ntl3) -- [photon, double] (nbl3),
  (ntr3) -- [photon, double] (nbr3),
  (nl3) -- [double] (n3) -- [double] (nr3),
  
  (ntl4) -- [photon, double] (nbl4),
  (ntr4) -- [photon, double] (nbr4),
  (nl4) -- [double] (n4) -- [double] (nr4),
};

\diagram* {
  (tt) -- [very thick] (bb),
};

\vertex [right=0.8cm of n1] {$\cdots$};
\vertex [right=0.8cm of n3] {$\cdots$};
\end{feynman}
\end{tikzpicture}
\caption{ \label{fig:1}}
\end{figure}

It is then straightforward to do multiple scattering of the dipole on the nucleus at impact parameter $b$. The results is \cite{cite8,cite9}
\begin{equation} \label{eq:2.2}
    T(b,x_\perp) = 1-S(b,x_\perp) = 1-e^{-Q_s^2 x_\perp^2/4}
\end{equation}
where
\begin{equation} \label{eq:2.3}
    Q_s^2(b) = \frac{4\pi^2\alpha C_F}{N_c^2-1} \rho L \,xG(x,x_\perp^2) \,,
\end{equation}
with $L$ the length of the nucleus at impact parameter $b$.
Noting that
\begin{equation} \label{eq:2.4}
    Q_s^2(b) \frac{x_\perp^2}{4} = T(x_\perp) \rho L \,,
\end{equation}
with the $T(x_\perp)$ as in \eqref{eq:2.1}, the exponentiation as given in \eqref{eq:2.2} is more or less evident in any gauge in which the scatterings in Fig.~\ref{fig:1} are sequential. Here we imagine $q = (q_+,q_-,q_\perp=0)$ where $q_+$ is large and $q_- = -Q^2/2q_+$, and we also imagine calculations in this projectile frame being done in an $A_+=0$ gauge.

One can go back to momentum space easily to get the leading jet distribution of flavor $f$ for coherent scatterings as
\begin{equation} \label{eq:2.5}
    \frac{\dd{\sigma_T^{\gamma^*}}}{\dd[2]{b} \dd[2]{p} \dd{z}} = 2\alpha_{em}N_c e_f^2[z^2+(1-z)^2] A
\end{equation}
with \cite{cite11}
\begin{equation} \label{eq:2.6}
    A = \bar{Q}^2 \int \frac{\dd[2]{x}\dd[2]{x'}}{(2\pi)^4} e^{i\underline{p}\vdot\qty(\underline{x}-\underline{x}')} K_1(\bar{Q}x_\perp) \frac{\underline{x}\vdot\underline{x}'}{x_\perp x'_\perp} K_1(\bar{Q}x'_\perp) T(b,x_\perp) T(b,x'_\perp)
\end{equation}
where $\bar{Q}^2 = Q^2 z(1-z)$.

\subsection{Projectile picture of inelastic reactions} \label{sec:2.2}

In case the nucleus breaks up during the scattering the inelastic cross section is given by
\begin{equation} \label{eq:2.7}
    \frac{\dd{\sigma_{in}}}{\dd[2]{b}} = 2\alpha_{em}N_c e_f^2 \bar{Q}^2 \int \frac{\dd[2]{x}}{4\pi^2} [z^2 + (1-z)^2] K_1^2(\bar{Q}x_\perp) [1-S^2(b,x_\perp)]\dd{z}
\end{equation}

When the scattering is strong, that is when $S(b,x_\perp) \simeq 0$ in \eqref{eq:2.7} and $T(b,x_\perp) \simeq T(b,x_\perp') \simeq 1$ in \eqref{eq:2.6} the elastic and inelastic cross sections are equal as is easily seen by integrating \eqref{eq:2.5} over $\dd[2]{p} \dd{z}$ which gives \eqref{eq:2.7}. Thus in the projectile picture scattering becomes strong, or equivalently the unitarity regime is reached, when the quark-antiquark separation of the dipole coming from the $\gamma^*$ is greater than $1/Q_s$. Now let us see how this corresponds to the idea of quark saturation.

\subsection{TMD's and quark saturation (Coherent scattering)} \label{sec:2.3}

Now suppose one does the calculation given in \eqref{eq:2.5} and \eqref{eq:2.6} and illustrated in Fig.~\ref{fig:1} in $A_-=0$ gauge. Several changes occur from the $A_+=0$ gauge calculation. First the scatterings are no longer 
sequential in $x_+$. Because the gluon propagator now has the form
\begin{equation} \label{eq:2.8}
    G_{\beta\alpha}(l) = \frac{-i}{l^2+i\epsilon} \qty[g_{\beta\alpha} - \frac{\eta_\beta l_\alpha}{l_--i\epsilon} - \frac{\eta_\alpha l_\beta}{l_-+i\epsilon}]
\end{equation}
the gluon interactions are very nonlocal in $x_+$. (The choice of $i\epsilon$'s in \eqref{eq:2.8} is not unique but the interactions are nonlocal for any choice of the $i\epsilon$ factors.) Thus the interactions of the dipole with the various nucleons in the nucleus are not sequential. Also, in $A_-=0$ gauge only the quark (or antiquark) with smaller longitudinal momentum $(q-p)_+ = (1-z)p_+$ or $p_+ = zp_+$ interacts with the nucleons in the nucleus. Thus the graph of Fig.~\ref{fig:1} takes the form shown in Fig.~\ref{fig:2A} \cite{cite13}, where we imagine the time variable of the Feynman graphs to be $x_+$ but the calculation is done in $A_-=0$ gauge. When we imagine the time variable to be $x_-$, thus 
matching the $A_-=0$ gauge so that the Feynman graphs also have a light cone perturbation theory interpretation, we draw the graphs as shown in Fig.~\ref{fig:2B}. We shall often refer to Fig.~\ref{fig:2A} as the projectile frame and Fig.~\ref{fig:2B} as the target frame. The $A_+=0$ gauge is natural for a projectile frame calculation while the $A_-=0$ gauge is natural for a target frame calculation. The parton picture becomes manifest in a target frame calculation using light cone perturbation theory with Fig.~\ref{fig:2B} showing the virtual photon hitting the struck quark, $p$, and thus measuring the sea quark distribution of the nucleus $A$.

\begin{figure}[htbp]
\centering
\begin{subfigure}[b]{0.99\textwidth}
\centering
\begin{equation*}
\begin{gathered}
\begin{tikzpicture}
\begin{feynman} 
\vertex (p1);
\vertex [right=1cm of p1] (l);
\vertex [below right=0.707cm of l] (b1);
\vertex [right=0.4cm of b1] (b2);
\vertex [right=1.0cm of b2] (b3);
\vertex [right=0.6cm of b3] (b4);
\vertex [right=0.6cm of b4] (b5);
\vertex [right=1.0cm of b5] (b6);
\vertex [right=0.4cm of b6] (b7);
\vertex [above right=0.707cm of b7] (r);
\vertex [right=1cm of r] (p2);
\vertex [above right=0.707cm of l] (t1);
\vertex [right=0.4cm of t1] (t2);
\vertex [right=1.0cm of t2] (t3);
\vertex [right=0.6cm of t3] (t4);
\vertex [right=0.6cm of t4] (t5);
\vertex [right=1.0cm of t5] (t6);
\vertex [right=0.4cm of t6] (t7);

\vertex [above=0.5cm of t4] (tt);
\vertex [below=0.5cm of b4] (bb);

\vertex [above=0.0cm of b2] (nt1);
\vertex [below=0.5cm of b2, blob, minimum height=0.5cm, minimum width=0.5cm] (n1) {};
\vertex [above=0.22cm of n1] (nb1);
\vertex [left=0.1cm of nt1] (ntl1);
\vertex [right=0.1cm of nt1] (ntr1);
\vertex [left=0.1cm of nb1] (nbl1);
\vertex [right=0.1cm of nb1] (nbr1);
\vertex [below left=0.5cm of n1] (nl1);
\vertex [below right=0.5cm of n1] (nr1);

\vertex [above=0.0cm of b3] (nt2);
\vertex [below=0.5cm of b3, blob, minimum height=0.5cm, minimum width=0.5cm] (n2) {};
\vertex [above=0.22cm of n2] (nb2);
\vertex [left=0.1cm of nt2] (ntl2);
\vertex [right=0.1cm of nt2] (ntr2);
\vertex [left=0.1cm of nb2] (nbl2);
\vertex [right=0.1cm of nb2] (nbr2);
\vertex [below left=0.5cm of n2] (nl2);
\vertex [below right=0.5cm of n2] (nr2);

\vertex [above=0.0cm of b5] (nt3);
\vertex [below=0.5cm of b5, blob, minimum height=0.5cm, minimum width=0.5cm] (n3) {};
\vertex [above=0.22cm of n3] (nb3);
\vertex [left=0.1cm of nt3] (ntl3);
\vertex [right=0.1cm of nt3] (ntr3);
\vertex [left=0.1cm of nb3] (nbl3);
\vertex [right=0.1cm of nb3] (nbr3);
\vertex [below left=0.5cm of n3] (nl3);
\vertex [below right=0.5cm of n3] (nr3);

\vertex [above=0.0cm of b6] (nt4);
\vertex [below=0.5cm of b6, blob, minimum height=0.5cm, minimum width=0.5cm] (n4) {};
\vertex [above=0.22cm of n4] (nb4);
\vertex [left=0.1cm of nt4] (ntl4);
\vertex [right=0.1cm of nt4] (ntr4);
\vertex [left=0.1cm of nb4] (nbl4);
\vertex [right=0.1cm of nb4] (nbr4);
\vertex [below left=0.5cm of n4] (nl4);
\vertex [below right=0.5cm of n4] (nr4);

\diagram* {
  (l) -- [quarter right] (b1) -- (b2) -- (b3) -- (b4) -- [fermion] (b5) -- (b6) -- (b7) -- [quarter right] (r),
  (l) -- [quarter left] (t1) -- (t2) -- (t3) -- [anti fermion] (t4) -- [edge label=\(q-p\), near end] (t5) -- (t7) -- [quarter left] (r),
  (p1) -- [charged boson, edge label'=\(q\)] (l),
  (r) -- [charged boson, edge label'=\(q\)] (p2),
  
  (ntl1) -- [photon, double] (nbl1),
  (ntr1) -- [photon, double] (nbr1),
  (nl1) -- [double] (n1) -- [double] (nr1),
  
  (b1) -- [photon, double,, in=90, out=-90] (nbl2),
  (ntr2) -- [photon, double] (nbr2),
  (nl2) -- [double] (n2) -- [double] (nr2),
  
  (ntl3) -- [photon, double] (nbl3),
  (ntr3) -- [photon, double] (nbl4),
  (nl3) -- [double] (n3) -- [double] (nr3),
  
  (ntl4) -- [photon, double] (nbr3),
  (ntr4) -- [photon, double] (nbr4),
  (nl4) -- [double] (n4) -- [double] (nr4),
};

\diagram* {
  (tt) -- [very thick] (bb),
};

\vertex [right=0.53cm of n1] {$\cdots$};
\vertex [right=0.53cm of n3] {$\cdots$};
\end{feynman}
\end{tikzpicture}
\end{gathered}
\equiv
\begin{gathered}
\begin{tikzpicture}
\begin{feynman} 
\vertex (p1);
\vertex [right=1cm of p1] (l);
\vertex [below right=0.707cm of l] (b1);
\vertex [right=0.25cm of b1] (b2);
\vertex [right=0.75cm of b2] (b3);
\vertex [right=0.75cm of b3] (b4);
\vertex [right=0.25cm of b4] (b5);
\vertex [above right=0.707cm of b5] (r);
\vertex [right=1cm of r] (p2);
\vertex [above right=0.707cm of l] (t1);
\vertex [right=0.25cm of t1] (t2);
\vertex [right=0.75cm of t2] (t3);
\vertex [right=0.75cm of t3] (t4);
\vertex [right=0.25cm of t4] (t5);

\vertex [above=0.5cm of t3] (tt);
\vertex [below=0.5cm of b3] (bb);

\vertex [above=0.3cm of b2] (nt1);
\vertex [below=0.5cm of b2, blob, minimum height=0.5cm, minimum width=0.5cm] (n1) {$A$};
\vertex [above=0.22cm of n1] (nb1);
\vertex [left=0.1cm of nt1] (ntl1);
\vertex [right=0.1cm of nt1] (ntr1);
\vertex [left=0.1cm of nb1] (nbl1);
\vertex [right=0.1cm of nb1] (nbr1);
\vertex [below left=0.5cm of n1] (nl1);
\vertex [below right=0.5cm of n1] (nr1);

\vertex [above=0.3cm of b4] (nt2);
\vertex [below=0.5cm of b4, blob, minimum height=0.5cm, minimum width=0.5cm] (n2) {$A$};
\vertex [above=0.22cm of n2] (nb2);
\vertex [left=0.1cm of nt2] (ntl2);
\vertex [right=0.1cm of nt2] (ntr2);
\vertex [left=0.1cm of nb2] (nbl2);
\vertex [right=0.1cm of nb2] (nbr2);
\vertex [below left=0.5cm of n2] (nl2);
\vertex [below right=0.5cm of n2] (nr2);

\diagram* {
  (l) -- [quarter right, fermion, edge label=$p$] (b1) -- (b2) -- (b3) -- [fermion, edge label=$\delta+p$, near end] (b4) -- (b5) -- [quarter right] (r),
  (l) -- [quarter left] (t1) -- (t2) -- [anti fermion] (t3) -- [edge label=\(q-p\), near end] (t4) -- (t5) -- [quarter left] (r),
  (p1) -- [charged boson, edge label'=\(q\)] (l),
  (r) -- [charged boson, edge label'=\(q\)] (p2),
  
  (b2) -- [gluon] (n1),
  (nl1) -- [double] (n1) -- [double] (nr1),
  
  (b4) -- [gluon] (n2),
  (nl2) -- [double] (n2) -- [double] (nr2),
};

\diagram* {
  (tt) -- [very thick] (bb),
};
\end{feynman}
\end{tikzpicture}
\end{gathered}
\end{equation*}
\caption{\label{fig:2A}}
\end{subfigure}
\hfill

\begin{subfigure}[b]{0.99\textwidth}
\centering
\begin{equation*}
\begin{gathered}
\begin{tikzpicture}
\begin{feynman} 
\vertex (t1);
\vertex [right=0.4cm of t1] (t2);
\vertex [right=1.2cm of t2] (t3);
\vertex [right=0.75cm of t3] (t4);
\vertex [right=0.75cm of t4] (t5);
\vertex [right=1.2cm of t5] (t6);
\vertex [right=0.4cm of t6] (t7);
\vertex [below=1.5cm of t3] (b1);
\vertex [below=1.5cm of t4] (b2);
\vertex [below=1.5cm of t5] (b3);
\vertex [below left=1cm of b1] (p1);
\vertex [below right=1cm of b3] (p2);

\vertex [above=0.5cm of t4] (tt);
\vertex [below=0.5cm of b2] (bb);

\vertex [below=0.0cm of t2] (nt1);
\vertex [above=0.5cm of t2, blob, minimum height=0.5cm, minimum width=0.5cm] (n1) {};
\vertex [below=0.22cm of n1] (nb1);
\vertex [left=0.1cm of nt1] (ntl1);
\vertex [right=0.1cm of nt1] (ntr1);
\vertex [left=0.1cm of nb1] (nbl1);
\vertex [right=0.1cm of nb1] (nbr1);
\vertex [above left=0.5cm of n1] (nl1);
\vertex [above right=0.5cm of n1] (nr1);

\vertex [below=0.0cm of t3] (nt2);
\vertex [above=0.5cm of t3, blob, minimum height=0.5cm, minimum width=0.5cm] (n2) {};
\vertex [below=0.22cm of n2] (nb2);
\vertex [left=0.1cm of nt2] (ntl2);
\vertex [right=0.1cm of nt2] (ntr2);
\vertex [left=0.1cm of nb2] (nbl2);
\vertex [right=0.1cm of nb2] (nbr2);
\vertex [above left=0.5cm of n2] (nl2);
\vertex [above right=0.5cm of n2] (nr2);

\vertex [below=0.0cm of t5] (nt3);
\vertex [above=0.5cm of t5, blob, minimum height=0.5cm, minimum width=0.5cm] (n3) {};
\vertex [below=0.22cm of n3] (nb3);
\vertex [left=0.1cm of nt3] (ntl3);
\vertex [right=0.1cm of nt3] (ntr3);
\vertex [left=0.1cm of nb3] (nbl3);
\vertex [right=0.1cm of nb3] (nbr3);
\vertex [above left=0.5cm of n3] (nl3);
\vertex [above right=0.5cm of n3] (nr3);

\vertex [below=0.0cm of t6] (nt4);
\vertex [above=0.5cm of t6, blob, minimum height=0.5cm, minimum width=0.5cm] (n4) {};
\vertex [below=0.22cm of n4] (nb4);
\vertex [left=0.1cm of nt4] (ntl4);
\vertex [right=0.1cm of nt4] (ntr4);
\vertex [left=0.1cm of nb4] (nbl4);
\vertex [right=0.1cm of nb4] (nbr4);
\vertex [above left=0.5cm of n4] (nl4);
\vertex [above right=0.5cm of n4] (nr4);

\diagram* {
  (t1) -- (t2) -- (t3) -- [anti fermion] (t4) -- [edge label'=\(\delta-p\), near end] (t5) -- (t7),
  (t1) -- [fermion, edge label'=$p$] (b1) -- (b2) -- [fermion, edge label=$q+p$, near end] (b3) -- (t7),
  (p1) -- [charged boson, edge label'=\(q\)] (b1),
  (b3) -- [charged boson, edge label'=\(q\)] (p2),
  
  (t1) -- [photon, double] (nbl1),
  (ntr1) -- [photon, double] (nbl2),
  (nl1) -- [double] (n1) -- [double] (nr1),
  
  (ntr2) -- [photon, double] (nbr2),
  (ntl2) -- [photon, double] (nbr1),
  (nl2) -- [double] (n2) -- [double] (nr2),
  
  (ntl3) -- [photon, double] (nbl3),
  (t7) -- [photon, double, in=310, out=90] (nbr3),
  (nl3) -- [double] (n3) -- [double] (nr3),
  
  (ntl4) -- [photon, double] (nbl4),
  (ntr4) -- [photon, double] (nbr4),
  (nl4) -- [double] (n4) -- [double] (nr4),
};

\diagram* {
  (tt) -- [very thick] (bb),
};

\vertex [right=0.63cm of n1] {$\cdots$};
\vertex [right=0.63cm of n3] {$\cdots$};
\end{feynman}
\end{tikzpicture}
\end{gathered}
\equiv
\begin{gathered}
\begin{tikzpicture}
\begin{feynman} 
\vertex (t1);
\vertex [right=1.5cm of t1] (t2);
\vertex [right=1.5cm of t2] (t3);
\vertex [below=1.0cm of t2] (b2);
\vertex [left=0.9cm of b2] (b1);
\vertex [right=0.9cm of b2] (b3);
\vertex [below left=1cm of b1] (p1);
\vertex [below right=1cm of b3] (p2);

\vertex [above=0.5cm of t2] (tt);
\vertex [below=0.5cm of b2] (bb);

\vertex [below] (nt1);
\vertex [above=0.5cm of t1, blob, minimum height=0.5cm, minimum width=0.5cm] (n1) {$A$};
\vertex [below=0.22cm of n1] (nb1);
\vertex [left=0.1cm of nt1] (ntl1);
\vertex [right=0.1cm of nt1] (ntr1);
\vertex [left=0.1cm of nb1] (nbl1);
\vertex [right=0.1cm of nb1] (nbr1);
\vertex [above left=1.0cm of n1] (nl1) {$P$};
\vertex [above right=1.0cm of n1] (nr1) {$P-\delta$};

\vertex [below=0.3cm of t3] (nt2);
\vertex [above=0.5cm of t3, blob, minimum height=0.5cm, minimum width=0.5cm] (n2) {$A$};
\vertex [below=0.22cm of n2] (nb2);
\vertex [left=0.1cm of nt2] (ntl2);
\vertex [right=0.1cm of nt2] (ntr2);
\vertex [left=0.1cm of nb2] (nbl2);
\vertex [right=0.1cm of nb2] (nbr2);
\vertex [above left=1.0cm of n2] (nl2) {$P-\delta$};
\vertex [above right=1.0cm of n2] (nr2) {$P$};

\diagram* {
  (t1) -- (t2) -- [anti fermion, edge label=$\delta+p$] (t3),
  (t1) -- [fermion, edge label'=$p$] (b1) -- (b2) -- [fermion, edge label=$q+p$] (b3) -- (t3),
  (p1) -- [charged boson, edge label'=\(q\)] (b1),
  (b3) -- [charged boson, edge label'=\(q\)] (p2),
  
  (t1) -- [gluon] (n1),
  (nl1) -- [double] (n1) -- [double] (nr1),
  
  (t3) -- [gluon] (n2),
  (nl2) -- [double] (n2) -- [double] (nr2),
};

\diagram* {
  (tt) -- [very thick] (bb),
};
\end{feynman}
\end{tikzpicture}
\end{gathered}
\end{equation*}
\caption{\label{fig:2B}}
\end{subfigure}
\caption{ \label{fig:2}}
\end{figure}

If one fixes $p_- = xP_-$ but integrates over $\underline{p}$ in Fig.~\ref{fig:2B} then the quark distribution of the 
nucleus is obtained. If one fixes both $p_- = xP_-$ and $\underline{p}$ then the transverse momentum dependent (TMD) quark distribution is obtained which we shall denote as
\begin{equation} \label{eq:2.9}
    \frac{\dd{}}{\dd[2]{b} \dd[2]{p}} \qty(x q_f(x,Q^2) + x\bar{q}_f(x,Q^2))
\end{equation}
if the impact parameter of the collision is fixed and as
\begin{equation} \label{eq:2.10}
    \frac{\dd{}}{\dd[2]{p}} \qty(x q_f(x,Q^2) + x\bar{q}_f(x,Q^2))
\end{equation}
if one integrates over all impact parameters. The TMD's depend on the Bjorken-$x$ variable, $Q^2$ and the transverse momentum of the struck quark, $\underline{p}$. (We shall always take the transverse momentum of the $\gamma^*$ to be zero.)

We can use \eqref{eq:2.5} and \eqref{eq:2.6} to get the coherent quark distribution by recalling that
\begin{equation} \label{eq:2.11}
    \sigma_T^{\gamma^*} = \frac{4\pi^2\alpha_{em}}{Q^2} F_2(x,Q^2) \,.
\end{equation}
Then one gets
\begin{multline} \label{eq:2.12}
    \frac{\dd{F_2^{f+\bar{f}}}}{\dd[2]{b} \dd[2]{p}} = \int_0^{1/2} \dd{z} \int \dd[2]{x}\dd[2]{x'} \frac{\bar{Q}^2 Q^2}{16\pi^6} e_f^2 N_c e^{i\underline{p}\vdot\qty(\underline{x}-\underline{x}')} \\
    K_1(\bar{Q}x_\perp) \frac{\underline{x}\vdot\underline{x}'}{x_\perp x'_\perp} K_1(\bar{Q}x'_\perp) T(b,x_\perp) T(b,x'_\perp) \,.
\end{multline}
We now include both quarks and antiquarks of flavor $f$. When $p_\perp \ll Q_s$, $T(b,x_\perp')$ and  $T(b,x_\perp)$ will be close to 1 and the integral in \eqref{eq:2.12} can be done to give \cite{cite9,cite13}
\begin{equation} \label{eq:2.13}
    \frac{\dd{F_2^{f+\bar{f}}}}{\dd[2]{b} \dd[2]{p}} = \frac{\dd{}}{\dd[2]{b} \dd[2]{p}} \qty(x q_f(x,Q^2) + x\bar{q}_f(x,Q^2))^{coherent} = \frac{N_c}{4\pi^4} \,.
\end{equation}

\eqref{eq:2.13} gives the (two dimensional) coherent occupation of quarks (and antiquarks) of flavor $f$ as $N_c$ times a pure number. Since \eqref{eq:2.13} refers to coherent nuclear reactions the quark and antiquark which are produced have equal and opposite transverse momentum $\underline{p}$ as is evident from the projectile frame graphs in Fig.~\ref{fig:2B}.

One can also show that the quark-antiquark pair produced in the coherent reaction and illustrated in Fig.~\ref{fig:2A} do not have additional quarks or gluons in the diffractively produced 
system. That is so long as $p_\perp/Q_s \ll 1$ the $\gamma^*$ diffractively, and coherently, only produces a single pair of $q\bar{q}$ ``jets''. In terms of the nuclear wavefunction, at the time the $\gamma^*$ hits the struck quark that wavefunction is a coherent state \cite{cite13} of zero momentum $q\bar{q}$ pairs which when freed from the nuclear wavefunction by the $\gamma^*$ are produced leaving the wavefunction exactly as it was before the $q\bar{q}$ pair are freed, that in the ground state of the nucleus.

It should now be apparent that in the coherent reactions we have been considering, unitarity in the $\gamma^*+A\to q\bar{q}+A$ reaction in the projectile picture of Fig.\ref{fig:2A} is equivalent to quark saturation of the light cone wavefunction of Fig.~\ref{fig:2B} where the $\gamma^*$ serves to measure the phase space density of the $q\bar{q}$ pairs in the wavefunction.

\subsection{TMD's and quark saturation (Inelastic scattering)} \label{sec:2.4}

The discussion of TMD's for inelastic reactions and the corresponding occupation for quarks in the saturation region follows the discussion we have already given for coherent reactions. The details are given in Ref.~\cite{cite13} and here we simply list the results for inelastic reactions corresponding to \eqref{eq:2.13} for coherent reactions. The result is
\begin{equation} \label{eq:2.14}
    \frac{\dd{}}{\dd[2]{b} \dd[2]{p}} \qty(x q_f + x\bar{q}_f)^{inelastic} = \frac{N_c}{4\pi^4}
\end{equation}
where $p_\perp^2 < Q_s^2$. We note that \eqref{eq:2.13} and \eqref{eq:2.14} are the same corresponding to the elastic dipole-nucleus cross section and the inelastic dipole-nucleus cross section being equal when the dipole size is greater than $1/Q_s$.

In the target picture of Fig.~\ref{fig:2B} for coherent reactions, and a similar picture for inelastic reactions, the parton densities of \eqref{eq:2.13} and \eqref{eq:2.14} are present in the wavefunction of the nucleus before the $\gamma^*$ hits the 
struck quark. In the case of inelastic reactions the struck quark (or antiquark) is paired with an antiquark (or quark) where, however, the total transverse momentum of the quark-antiquark pair is not zero and the color charge of the pair is also typically not zero. The $\gamma^*$ hitting the struck quark destroys the coherence of the nuclear wavefunction and the nucleus breaks up. In the case of coherent reactions the struck quark (or antiquark) is paired with an antiquark (or quark) with the pair having zero color charge and transverse momentum. In this case when the $\gamma^*$ hits the struck quark the quark-antiquark pair is produced leaving the nucleus in its ground state. This must happen because in the projectile picture that is exactly what emerges from a coherent strong interaction, $\gamma^*+A \to(q\bar{q})+A$. In the target picture this means that the nuclear wavefunction is a coherent 
state of zero momentum and zero color charge $q\bar{q}$ pairs. This picture is essentially manifest in the projectile picture but remains mysterious in the target picture. It is a great challenge to understand how QCD dynamics manages to arrange the light cone wavefunction of a nucleus (or hadron) in such a way that it is a coherent state of $q\bar{q}$ pairs having zero color charge and zero total transverse momentum.

In the pictures in Fig.~\ref{fig:1} and Fig.~\ref{fig:2} that we have used a simple picture of the nucleus, with no QCD evolution, has been used. All our results are independent of this choice. When we turn to gluon saturation we shall explicitly also allow QCD evolution of the nucleus. Finally, a word about nuclear shadowing and its relation to saturation. Formula \eqref{eq:2.13} is a formula 
reflecting saturation and it follows from the unitarity limit of the $q\bar{q}$ dipole scattering on the nucleus. This unitarity limit also gives strong nuclear shadowing. (Recall, without shadowing the left-hand side of \eqref{eq:2.13} would be proportional to $A^{1/3}$.) Unitarity corrections to the Born approximation give nuclear shadowing. In the literature one often distinguishes leading twist from higher twist shadowing \cite{cite10}. The shadowing expressed in \eqref{eq:2.13} and in Fig.~\ref{fig:2} is leading twist shadowing. This is evident from the target frame picture of Fig.~\ref{fig:2B} where it is clear that the struck quark does not interact after absorbing the $\gamma^*$, and this is also the criterion of a leading twist calculation when using $A_-=0$ gauge.

\section{Gluon saturation} \label{sec:3}

We turn now to TMD's for gluons and to gluon saturation in QCD. We don't have an external current which couples directly to gluons, as did the $\gamma^*$ to quarks in the previous section. Nevertheless it is possible to create an effective gluon dipole out of a compact (high relative transverse momentum) $q(\underline{p}) \bar{q}(-\underline{p}-\underline{k})$ pair along with a softer gluon $g(\underline{k})$. The analog of Fig.~\ref{fig:1} for a quark dipole scattering on a nucleus is now shown in Fig.~\ref{fig:3} for an (effective) gluon dipole scattering on a nucleus. The $q\bar{q}$ pair has a transverse size which is on the order of $1/p_\perp$ which we shall take to be the inverse of the $\gamma^*$ virtuality $Q$, while $k_\perp \ll Q$ and we shall often take $k_\perp \ll Q_s$ in what 
follows. The $q\bar{q}$ pair, after the gluon $g(k)$ has been emitted, looks like a gluon except if probed at a scale of $Q$ or greater. Thus the scattering depicted in Fig.~\ref{fig:3} is effectively that of a gluon dipole on an MV nucleus just as the scattering depicted in Fig.~\ref{fig:1} is that of a quark dipole scattering on an MV nucleus \cite{cite11,cite12}.

\begin{figure}[htbp]
\centering
\begin{tikzpicture}
\begin{feynman} 
\vertex (p1) {$\gamma*$};
\vertex [right=1.5cm of p1] (qa1);
\vertex [above right=0.35cm of qa1] (q1);
\vertex [below right=0.35cm of qa1] (a1);
\vertex [right=0.3cm of qa1] (t1);
\vertex [below=0.5cm of t1] (ml);
\vertex [below right=1.2cm of ml] (b1);
\vertex [right=0.1cm of b1] (b2);
\vertex [right=1.2cm of b2] (b3);
\vertex [right=1.5cm of b3] (b4);
\vertex [right=0.75cm of b3] (bm);
\vertex [right=0.2cm of bm] (bmr);
\vertex [above=2.2cm of bm] (tt);
\vertex [below=0.2cm of bm] (bb);
\vertex [right=1.2cm of b4] (b5);
\vertex [right=0.1cm of b5] (b6);
\vertex [above right=1.2cm of b6] (mr);
\vertex [above=0.5cm of mr] (t2);
\vertex [right=0.3cm of t2] (qa2);
\vertex [above left=0.35cm of qa2] (q2);
\vertex [below left=0.35cm of qa2] (a2);
\vertex [right=1.5cm of qa2] (p2) {$\gamma*$};

\vertex [above=0.25cm of b2] (nt1);
\vertex [below=0.25cm of b2, blob, minimum height=0.5cm, minimum width=0.5cm] (n1) {};
\vertex [above=0.22cm of n1] (nb1);
\vertex [left=0.1cm of nt1] (ntl1);
\vertex [right=0.1cm of nt1] (ntr1);
\vertex [left=0.1cm of nb1] (nbl1);
\vertex [right=0.1cm of nb1] (nbr1);
\vertex [below left=0.5cm of n1] (nl1);
\vertex [below right=0.5cm of n1] (nr1);
\vertex [above=0.25cm of b3] (nt2);
\vertex [below=0.25cm of b3, blob, minimum height=0.5cm, minimum width=0.5cm] (n2) {};
\vertex [above=0.22cm of n2] (nb2);
\vertex [left=0.1cm of nt2] (ntl2);
\vertex [right=0.1cm of nt2] (ntr2);
\vertex [left=0.1cm of nb2] (nbl2);
\vertex [right=0.1cm of nb2] (nbr2);
\vertex [below left=0.5cm of n2] (nl2);
\vertex [below right=0.5cm of n2] (nr2);
\vertex [above=0.25cm of b4] (nt3);
\vertex [below=0.25cm of b4, blob, minimum height=0.5cm, minimum width=0.5cm] (n3) {};
\vertex [above=0.22cm of n3] (nb3);
\vertex [left=0.1cm of nt3] (ntl3);
\vertex [right=0.1cm of nt3] (ntr3);
\vertex [left=0.1cm of nb3] (nbl3);
\vertex [right=0.1cm of nb3] (nbr3);
\vertex [below left=0.5cm of n3] (nl3);
\vertex [below right=0.5cm of n3] (nr3);
\vertex [above=0.25cm of b5] (nt4);
\vertex [below=0.25cm of b5, blob, minimum height=0.5cm, minimum width=0.5cm] (n4) {};
\vertex [above=0.22cm of n4] (nb4);
\vertex [left=0.1cm of nt4] (ntl4);
\vertex [right=0.1cm of nt4] (ntr4);
\vertex [left=0.1cm of nb4] (nbl4);
\vertex [right=0.1cm of nb4] (nbr4);
\vertex [below left=0.5cm of n4] (nl4);
\vertex [below right=0.5cm of n4] (nr4);
\vertex [right=2.8cm of q1] (qm);
\vertex [left=0.75cm of qm] (qml);
\vertex [right=2.8cm of a1] (am);
\vertex [left=0.75cm of am] (aml);

\diagram*{
  (p1)  -- [charged boson, edge label=$q$] (qa1) -- [quarter left] (q1) -- (qml) -- [fermion, edge label=$\underline{p}$, near end] (qm) -- (q2) -- [quarter left] (qa2) -- [charged boson, edge label=$q$] (p2),
  (qa1) -- [quarter right] (a1) -- (aml) -- [anti fermion, edge label'=$-\underline{p}-\underline{k}$, near start] (am) -- (a2) -- [quarter right] (qa2),
  (t1) -- [charged boson, quarter right, edge label'=$k$] (b1) -- [photon] (b3) -- [charged boson, edge label'=$\delta+k$, near end] (bmr) -- [photon] (b6) -- [charged boson, quarter right, edge label'=$k$] (t2),
  (ntl1) -- [photon, double] (nbl1),
  (ntr1) -- [photon, double] (nbr1),
  (nl1) -- [double] (n1) -- [double] (nr1),
  (ntl2) -- [photon, double] (nbl2),
  (ntr2) -- [photon, double] (nbr2),
  (nl2) -- [double] (n2) -- [double] (nr2),
  (ntl3) -- [photon, double] (nbl3),
  (ntr3) -- [photon, double] (nbr3),
  (nl3) -- [double] (n3) -- [double] (nr3),
  (ntl4) -- [photon, double] (nbl4),
  (ntr4) -- [photon, double] (nbr4),
  (nl4) -- [double] (n4) -- [double] (nr4),
};

\diagram* {
  (tt) -- [very thick] (bb),
};

\vertex [right=0.63cm of n1] {$\cdots$};
\vertex [right=0.63cm of n3] {$\cdots$};
\end{feynman}
\end{tikzpicture}
\caption{ \label{fig:3}}
\end{figure}

\subsection{Semiclassical gluons} \label{sec:3.1}

We shall begin our discussion of gluon saturation with a much simpler process which we start by viewing in the projectile frame where we now take a ``current''
\begin{equation} \label{eq:3.1}
    j(x) = -\frac{1}{4} F_{\mu\nu}^i F_{\mu\nu}^i
\end{equation}
which produces a high energy gluon somewhere in a MV nucleus and then that gluon multiply scatters as it passes through the nucleus as illustrated in Fig.~\ref{fig:4}. The multiple scatterings 
in the amplitude and complex conjugate amplitude lead to a gluon dipole scattering in the nucleus and, with an integration over the longitudinal position of the first scattering, gives the number of produced gluons as \cite{cite8, cite23}

\begin{figure}[htbp]
\centering
\begin{tikzpicture}
\begin{feynman} 
\vertex (q);
\vertex [below right=1.5cm of q] (t1);
\vertex [right=1.5cm of t1] (t2);
\vertex [right=1.5cm of t2] (t3);
\vertex [right=1.5cm of t3] (t4);
\vertex [right=1.5cm of t4] (t5);
\vertex [below=1.5cm of t1, blob, minimum height=0.5cm, minimum width=0.5cm] (b1) {};
\vertex [below=1.5cm of t2, blob, minimum height=0.5cm, minimum width=0.5cm] (b2) {};
\vertex [below=1.5cm of t4, blob, minimum height=0.5cm, minimum width=0.5cm] (b4) {};
\vertex [below left=0.9cm of b1] (l1) {$P$};
\vertex [right=0.5cm of b1] (r1);
\vertex [above=0.2cm of r1] (tr1);
\vertex [below=0.2cm of r1] (br1);
\vertex [below left=0.5cm of b2] (l2);
\vertex [right=0.5cm of b2] (r2);
\vertex [above=0.2cm of r2] (tr2);
\vertex [below=0.2cm of r2] (br2);
\vertex [below left=0.5cm of b4] (l4);
\vertex [right=0.5cm of b4] (r4);
\vertex [above=0.2cm of r4] (tr4);
\vertex [below=0.2cm of r4] (br4);

\diagram* {
  (q) -- [charged scalar, edge label=$q$] (t1) -- [charged boson, edge label=$q+\delta_1$] (t2) -- [charged boson, edge label=$q+\delta_2$] (t3) -- [photon] (t4) -- [charged boson, edge label=$q+k$] (t5),
  (b1) -- [charged boson, edge label'=$\delta_1$] (t1),
  (b2) -- [charged boson, edge label'=$\delta_2-\delta_1$] (t2),(b4) -- [charged boson, edge label'=$k-\delta_{n-1}$] (t4),
  (l1) -- [double] (b1) -- (r1),
  (tr1) -- (b1) -- (br1),
  (l2) -- [double] (b2) -- (r2),
  (tr2) -- (b2) -- (br2),
  (l4) -- [double] (b4) -- (r4),
  (tr4) -- (b4) -- (br4),
};

\vertex [below left=0.8cm of t4] {$\cdots$};
\vertex [below=0.6cm of b1] {$1$};
\vertex [below=0.6cm of b2] {$2$};
\vertex [below=0.6cm of b4] {$n$};
\end{feynman}
\end{tikzpicture}
\caption{ \label{fig:4}}
\end{figure}

\begin{equation} \label{eq:3.2}
    \frac{\dd{N}}{\dd[2]{b}\dd[2]{k}} = \int \frac{\dd[2]{x_\perp}}{(2\pi)^2} e^{i \underline{k} \vdot \underline{x}} \tilde{N}(\underline{x},\underline{b})
\end{equation}
where
\begin{equation} \label{eq:3.3}
    \tilde{N}(\underline{x},\underline{b}) = \frac{N_c^2-1}{\pi^2\alpha N_c \underline{x}^2} \qty(1-e^{-\underline{x}^2 Q_s^2/4})
\end{equation}
and where
\begin{equation} \label{eq:3.4}
    Q_s^2 = \frac{8\pi^2\alpha N_c}{N_c^2-1} \sqrt{R^2-b^2} \rho\, xG
\end{equation}
with $\rho$ the nuclear density and $xG$ the gluon distribution of a nucleon in the nucleus. The gluon in the amplitude can be imagined to be produced at $\underline{x}=0$ while the gluon in the complex conjugate amplitude produced at $\underline{x}$ so that the $e^{-\underline{x}^2 Q_s^2/4}$ in \eqref{eq:3.3} becomes the 
elastic gluon dipole scattering amplitude \cite{cite23} in the nucleus. Because the initial gluon production by the current $j$ is inelastic the distribution given by \eqref{eq:3.2} is the inelastic gluon phase space density which in this frame are produced gluons. When $Q_s^2$ is large \eqref{eq:3.2} can be evaluated in a logarithmic approximation to be
\begin{equation} \label{eq:3.5}
    \frac{\dd{N}}{\dd[2]{b} \dd[2]{k}} \simeq \frac{N_c^2 - 1}{4\pi^3 \alpha N_c} \int_{1/Q_s^2}^{1/k_\perp^2} \frac{\dd{x_\perp^2}}{x_\perp^2} = \frac{N_c^2-1}{4\pi^3\alpha N_c} \ln\frac{Q_s^2}{k_\perp^2} \,.
\end{equation}

To interpret \eqref{eq:3.5} in terms of partons in a light cone wavefunction it is better to view the calculation in the frame where the target nucleons have large momentum, $P_-$, and the calculation is done in an $A_-=0$ gauge. The picture then is illustrated in Fig.~\ref{fig:5}. (Light cone gauges are boost invariant so the value of $P_-$ is not important, however, a large value 
of $P_-$ makes the view of $x_-$ as the time variable more natural and the picture of the process in Fig.~\ref{fig:5} more intuitive.)

\begin{figure}[htbp]
\centering
\begin{tikzpicture}
\begin{feynman} 
\vertex [blob, minimum height=0.5cm, minimum width=0.5cm] (n1) {};
\vertex [left=0.8cm of n1] (l1) {$P$};
\vertex [right=0.5cm of n1] (r1);
\vertex [above=0.2cm of r1] (tr1);
\vertex [below=0.2cm of r1] (br1);
\vertex [right=3cm of n1, blob, minimum height=0.5cm, minimum width=0.5cm] (n2) {};
\vertex [left=0.5cm of n2] (l2);
\vertex [right=0.5cm of n2] (r2);
\vertex [above=0.2cm of r2] (tr2);
\vertex [below=0.2cm of r2] (br2);
\vertex [right=3cm of n2, blob, minimum height=0.5cm, minimum width=0.5cm] (n3) {};
\vertex [left=0.5cm of n3] (l3);
\vertex [right=0.5cm of n3] (r3);
\vertex [above=0.2cm of r3] (tr3);
\vertex [below=0.2cm of r3] (br3);
\vertex [below right=2.2cm of n1] (b1);
\vertex [below left=1.12cm of n2] (t1);
\vertex [below left=1.12cm of n3] (t2);
\vertex [right=3cm of b1] (b2);
\vertex [right=1.5cm of b2] (b3);
\vertex [below left=1.5cm of b2] (q);

\diagram* {
  (l1) -- [double] (n1) -- (r1),
  (tr1) -- (n1) -- (br1),
  (l2) -- [double] (n2) -- (r2),
  (tr2) -- (n2) -- (br2),
  (l3) -- [double] (n3) -- (r3),
  (tr3) -- (n3) -- (br3),
  (n1) -- [photon, quarter right] (b1) -- [charged boson, edge label'=$k$] (b2) -- [charged boson, edge label'=$q+k$] (b3),
  (b1) -- [photon] (t1) -- [photon] (n2),
  (t1) -- [photon] (t2) -- [photon, quarter right] (n3),
  (q) -- [charged scalar, edge label'=$q$] (b2),
};

\vertex [above left=0.5cm of t2] {$\cdots$};
\end{feynman}
\end{tikzpicture}
\caption{ \label{fig:5}}
\end{figure}

In Fig.~\ref{fig:5} a virtual gluon coming from any nucleon in the nucleus is then gauge rotated by nucleons at a similar impact parameter and having larger values of $x_-$ before it is struck by the current, $q$, and produced. In this picture the produced gluon is part of the light cone wavefunction of the nucleus until the current $q$ hits and produces it. The various gauge rotations of Fig.~\ref{fig:5} correspond to the multiple scatterings of Fig.~\ref{fig:4}. There is no nuclear shadowing in the simple model. The total number of gluons produced at a given impact parameter is proportional to the length $L\sim A^{1/3}$ 
of the nucleus at the impact parameter in question.

The growth of the produced gluon phase space density in \eqref{eq:3.5} or, equivalently, the growth of the phase space occupancy in the light cone wavefunction with the size of the nucleus is due to the inability of the multiple scatterings (the gauge rotations) to keep up with the number of gluons produced. When $Q_s^2$ is very large corresponding to a very large nucleus the number of gluons at a small $k_\perp \ll Q_s$ in the light cone wavefunction can become much larger than $1/\alpha$ as is evident from \eqref{eq:3.5}. In fact from  \eqref{eq:3.5} the gluon occupancy can become arbitrarily large as the nucleus become large. In this simple model the gluons in the light cone wavefunction only interact through gauge rotations by other gluons in the light cone wavefunction and those interactions, corresponding 
to the multiple scatterings in the projectile picture, are not sufficient to limit gluon occupancies. As we shall see in detail below this appears to be a general property of gluon saturation.

Since the growth of the gluon TMD given in \eqref{eq:3.5}, $\frac{1}{\alpha}\ln\frac{Q_s^2}{k_\perp^2}$, is going to be a property of all the gluon TMD's which we are going to discuss it is perhaps useful to go back a bit to see exactly how the TMD in \eqref{eq:3.5} comes about. One can write \eqref{eq:3.2} more completely as
\begin{equation} \label{eq:3.6}
    \frac{\dd{N}}{\dd[2]{b} \dd[2]{k}} = \frac{(N_c^2 - 1)\hat{q}}{(2\pi)^2 \alpha N_c} \int \frac{\dd[2]{x}}{(2\pi)^2} e^{i\underline{k}\vdot\underline{x}} \int_0^L\dd{z} e^{-\frac{\hat{q} x_\perp^2}{4} (L-z)}
\end{equation}

\begin{figure}[htbp]
\centering
\begin{tikzpicture}
\begin{feynman}
\vertex (t);
\vertex [below right=4cm of t] (r);
\vertex [below left=4cm of r] (b);
\vertex [below left=4cm of t] (l);
\vertex [below=2.8284271247cm of t, dot] (c) {};
\vertex [above right=2.8284271247cm of c] (tr);
\vertex [below=1cm of c] (line_anchor);

\vertex [left=1cm of line_anchor, dot] (z_vert) {};
\vertex [left=1.55cm of z_vert] (line_l);
\vertex [right=3.644cm of z_vert] (line_r);
\vertex [right=1.5cm of line_r] (line_rr);
\vertex [above left=1.5cm of z_vert] (q);

\vertex [below=0.75cm of z_vert, blob, minimum height=0.5cm, minimum width=0.5cm] (n1) {};
\vertex [below left=0.5cm of n1] (nl1);
\vertex [right=0.5cm of n1] (nr1);
\vertex [above=0.2cm of nr1] (ntr1);
\vertex [below=0.2cm of nr1] (nbr1);

\vertex [right=2cm of n1, blob, minimum height=0.5cm, minimum width=0.5cm] (n2) {};
\vertex [right=2cm of z_vert] (n2_vert);
\vertex [below left=0.5cm of n2] (nl2);
\vertex [right=0.5cm of n2] (nr2);
\vertex [above=0.2cm of nr2] (ntr2);
\vertex [below=0.2cm of nr2] (nbr2);

\vertex [above right=1.4142135624 of c] (Rlab_tmp);
\vertex [above left=0.05cm of Rlab_tmp] {$R$};

\vertex [right=0.1cm of c] (blab_t);
\vertex [above right=0.14cm of line_anchor] (blab_b);

\vertex [below right=0.1cm of line_l] (zlab_l);
\vertex [below left=0.1cm of z_vert] (zlab_r);

\vertex [above right=0.1cm of z_vert] (Lmzlab_l);
\vertex [above left=0.1cm of line_r] (Lmzlab_r);

\diagram* {
  (t) -- [quarter left] (r) -- [quarter left] (b) -- [quarter left] (l) -- [quarter left] (t),
  (c) -- (tr),
  (line_l) -- (z_vert),
  (z_vert) -- [photon] (line_r) -- [charged boson, edge label=$\underline{x}$, edge label'=$\underline{k}$] (line_rr),
  (q) -- [charged scalar, edge label=$q$] (z_vert),
  (z_vert) -- [photon] (n1),
  (nl1) -- [double] (n1) -- (nr1),
  (ntr1) -- (n1) -- (nbr1),
  (n2_vert) -- [photon] (n2),
  (nl2) -- [double] (n2) -- (nr2),
  (ntr2) -- (n2) -- (nbr2),
};

\draw [decoration={brace}, decorate] (blab_t.east) -- (blab_b.east)
node [pos=0.5, right] {\(b\)};
\draw [decoration={brace}, decorate] (zlab_r.south) -- (zlab_l.south)
node [pos=0.5, below] {\(z\)};
\draw [decoration={brace}, decorate] (Lmzlab_l.north) -- (Lmzlab_r.north)
node [pos=0.5, above, near end] {\(L-z=z'\)};

\vertex [right=1cm of n1] {$\cdots$};
\end{feynman}
\end{tikzpicture}
\caption{ \label{fig:6}}
\end{figure}

where the kinematics are shown in Fig.~\ref{fig:6} where the current enters the nucleus at impact parameter $b$ going in the $z$-direction. The length of nuclear material that the current encounters is
\begin{equation} \label{eq:3.7}
    L = 2\sqrt{R^2-b^2} \,.
\end{equation}
Eq.~\eqref{eq:3.6} corresponds to the current striking a nucleon at a distance $z$ from the back of the nucleus, producing a gluon at $z$ which then passes a length $(L-z)$ of the nuclear material before exiting the nucleus and giving the gluon TMD of the nucleus as in \eqref{eq:3.6}.

Now let's do the $\dd[2]{x}$ integral in \eqref{eq:3.6}. One easily finds
\begin{equation} \label{eq:3.8}
    \frac{\dd{N}}{\dd[2]{b} \dd[2]{k}} = \frac{N_c^2 - 1}{4\pi^3} \frac{1}{\alpha N_c} \int_0^L \frac{\dd{z}}{L-z} e^{-\frac{\underline{k}^2}{(L-z) \hat{q}}} \,.
\end{equation}
Call $L-z = z'$ so that
\begin{equation} \label{eq:3.9}
    \frac{\dd{N}}{\dd[2]{b} \dd[2]{k}} = \frac{N_c^2 - 1}{4\pi^3 \alpha N_c} \int_0^L \frac{\dd{z'}}{z'} e^{-\frac{\underline{k}^2 L}{Q_s^2 z'}}
\end{equation}
and \eqref{eq:3.9} gives the previous answer, \eqref{eq:3.5}. However, now from \eqref{eq:3.9} it is more transparent how the result \eqref{eq:3.5} comes about. For example when $Q_s^2/k_\perp^2$ is of order 1 it is clear from \eqref{eq:3.9} that the result comes from $z'\sim L$ that is the produced gluon must travel over a length 
$L$ in the nucleus for multiple scattering to build up a transverse momentum of size $Q_s$. On the other hand when $k_\perp^2/Q_s^2 \ll 1$ all $z'$ values between $\frac{k_\perp^2}{Q^2} L$ and $L$ contribute. Here one can have $z'$ as small as $\frac{k_\perp^2}{Q^2} L$ and build a final transverse momentum for gluon $k$ by multiple scattering. However, one can also have $z'$ as big as $L$ and also get contributions to a final gluon having $k_\perp$. The relative weighting of the different regions of $z'$ is given by the $\frac{\dd{z'}}{z'}$ integration in \eqref{eq:3.8}. What is interesting is that when $z' \gg \frac{k_\perp^2}{Q^2} L$ and there is a lot of multiple scattering there are still a significant number of gluons at $k_\perp^2$ being produced. Thus the $\ln Q_s^2/k_\perp^2$ which appears in \eqref{eq:3.5} is given by rather large distances, $z'$, where the typical gluon transverse momentum is $Q_s^2 z'/L \gg k_\perp^2$ but where there is still a 
significant probability of having gluons at $k_\perp$. As we shall see below more realistic pictures of the gluon TMD of a large nucleus at high energy exhibit a picture much like that we have just discussed. The light cone wavefunction of the nucleus has gluons at $Q_s$ with occupation $1/\alpha$ and these are the large majority of gluons in the light cone wavefunction. In addition there are a smaller number of gluons at small $k_\perp$, with $k_\perp/Q_s \ll 1$, but with higher occupancy and the size of the occupation of these gluons can be arbitrarily large until Sudakov effects enter to limit the occupancy.

Finally in this quasiclassical (tree graph) approximation there are no coherent reactions so that the TMD in \eqref{eq:3.2} is only for inelastic reactions. To get a coherent reaction one could 
let the ``current'' $j$ in \eqref{eq:3.1} first split into a $g(\underline{k}) g(-\underline{k})$ pair and then have this gluon dipole scatter on the nucleus. This would go well beyond the semiclassical gluon approximation and is closer in spirit to the $q\bar{q}g$ scattering which we describe below. Thus our quasiclassical gluon approximation only involves inelastic reactions but does have very large occupancies, for gluons in the light cone wavefunctions having $k_\perp/Q_s \ll 1$, as given by \eqref{eq:3.5}.

We close this section on the quasiclassical gluon approximation by noting that there is no gluon shadowing in this approximation. This is not surprising since one expects shadowing only in a full quantum treatment of gluon production. The lack of shadowing is evident from \eqref{eq:3.2}.
Integrating \eqref{eq:3.2} over $k_\perp$ gives
\begin{equation} \label{eq:3.10}
    xG_A = N = \int\dd[2]{b}\dd[2]{x_\perp} \frac{\dd{N}}{\dd[2]{b}\dd[2]{k_\perp}} = \int\dd[2]{b} \tilde{N}(x_\perp=0,\underline{b})
\end{equation}
or, using \eqref{eq:3.4}
\begin{equation} \label{eq:3.11}
    xG_A = \frac{N_c^2-1}{4\pi^2\alpha N_c} \int Q_s^2 \dd[2]{b} = A\, xG
\end{equation}
giving the nuclear gluon distribution as the atomic number times the nucleon gluon distribution.

\subsection{Gluon dipole scattering on an MV nucleus, coherent scattering} \label{sec:3.2}

In this section we introduce the idea of a gluon dipole in detail. Our picture of quark saturation was helped by the natural quark dipole nucleus scattering which appears in electron scattering on a nucleus. Here, again, we consider electron-nucleus scattering, and to begin we work in a projectile frame 
where the incoming $\gamma^*$ first breaks up into a quark-antiquark pair. We consider the case where the quark and antiquark each have a (large) transverse momentum about equal to the virtuality, $Q$, of the $\gamma^*$. When $Q$ is large such a quark-antiquark dipole interacts very weakly with the nucleus. To make the scattering with the nucleus strong we take events where the forward moving system is a $(q\bar{q}g)$ and where the gluon has transverse momentum $k_\perp$ with $k_\perp < Q_s$. We are thus looking at coherent diffraction where the nucleus does not break up during the scattering and a quark, an antiquark and a gluon jet are produced with the quark and antiquark jets having large transverse momenta. After emitting 
the gluon before striking the nucleus the quark-antiquark pair interacts with the nucleus as a gluon, so that the scattering is that of a gluon dipole with each gluon having transverse momentum $k_\perp$. The coherent scattering process is pictured in Fig.~\ref{fig:3}. We can pretty much guess the form of the coherent gluon TMD, $\frac{\dd{N}}{\dd[2]{b}\dd[2]{k}}$, from the similarity of the picture of the gluon process in Fig.~\ref{fig:3} compared to that of the quark process in Fig.~\ref{fig:1} or in Fig.~\ref{fig:2B}. Indeed, if we were to calculate the graphs of Fig.~\ref{fig:3} in an $A_-=0$ light cone gauge and picture the process with $x_-$ as the time variable the coherent scattering would look like that of Fig.~\ref{fig:7} which is to be compared to Fig.~\ref{fig:2B}. For the coherent gluon TMD what replaces \eqref{eq:2.13} is 

\begin{figure}[htbp]
\centering
\begin{tikzpicture}
\begin{feynman} 
\vertex [blob, minimum height=0.5cm, minimum width=0.5cm] (n1) {$A$};
\vertex [left=0.8cm of n1] (nl1) {$P$};
\vertex [right=1.1cm of n1] (nr1) {$P-\delta$};
\vertex [below=1.5cm of n1] (tl);
\vertex [below right=1.5cm of tl] (sul);
\vertex [right=0.5cm of sul] (sur);
\vertex [above right=1.5cm of sur] (tr);
\vertex [right=2.62cm of n1, blob, minimum height=0.5cm, minimum width=0.5cm] (n2) {$A$};
\vertex [left=0.5cm of n2] (nl2);
\vertex [right=0.5cm of n2] (nr2);
\vertex [below=0.5cm of sul] (sbl);
\vertex [right=0.5cm of sbl] (sbr);
\vertex [below left=1cm of sbl] (p1);
\vertex [below right=1cm of sbr] (p2);
\vertex [right=0.25cm of sul] (sm);
\vertex [above=1.5cm of sm] (tt);
\vertex [above=1.05cm of sm] (tm);
\vertex [below=1.0cm of sm] (bb);

\diagram* {
  (nl1) -- [double] (n1) -- [double] (nr1),
  (nl2) -- [double] (n2) -- [double] (nr2),
  (n1) -- [gluon] (tl) -- [photon] (tm) -- [charged boson, edge label=$\delta-k$] (tr) -- [gluon] (n2),
  (tl) -- [charged boson, edge label'=$k$] (sul) -- (sur) -- [photon] (tr),
  (sul) -- (sbl) -- (sbr) -- (sur),
  (p1) -- [charged boson, edge label=$q$] (sbl),
  (sbr) -- [charged boson, edge label=$q$] (p2),
};

\diagram* {
  (tt) -- [very thick] (bb),
};
\end{feynman}
\end{tikzpicture}
\caption{ \label{fig:7}}
\end{figure}

\begin{equation} \label{eq:3.12}
    \frac{\dd{N}}{\dd[2]{b}\dd[2]{k}} = \frac{\dd{xG_A}}{\dd[2]{b}\dd[2]{k}} = \frac{N_c^2-1}{8\pi^4} (Y-Y_0) \,.
\end{equation}
While \eqref{eq:3.12} is derived in detail in Refs.~\cite{cite9,cite14,cite16} we can give a very heuristic argument for \eqref{eq:3.12} from \eqref{eq:2.13}. The factors for the coherent quark TMD given in \eqref{eq:2.13} are
\begin{equation*}
    \frac{\dd{F_2^{f+\bar{f}}}}{\dd[2]{b} \dd[2]{p}} = \frac{1}{(2\pi)^4} \cdot N_c \cdot 2 \cdot 2
\end{equation*}
where the $N_c$ counts the number of colors of quarks and the two factors of 2 count the number of spin components and the fact that one counts both quarks and antiquarks. Going to the coherent gluon TMD it is pretty clear that the $N_c$ in \eqref{eq:2.13} should become $N_c^2-1$, the number of colors of gluons, the factor of 2 counting spins should remain the same while the factor of 2 counting quarks and antiquarks should be absent. After these changes the difference between \eqref{eq:3.12} and \eqref{eq:2.13} is just the factor $(Y-Y_0)$. The factor $Y-Y_0$ naturally 
appears in \eqref{eq:3.12} because the $\frac{\dd{k_+}}{k_+}$ integral in the graphs of Fig.~\ref{fig:3} has a lower limit of integration determined by $2k_+/k_\perp^2$ being greater than $R$ the nuclear size which gives $Y_0$ while $Y$ is given by the upper limit at the $\dd{k_+}$-integration being given by the requirement that $2k_+/k_\perp^2 < 2q_+/Q^2$ so that
\begin{equation} \label{eq:3.13}
    \int_R^{2q_+/Q^2} \frac{\dd{(2k_+/k_\perp^2)}}{(2k_+/k_\perp^2)} = \ln\frac{2q_+}{R Q^2} = \ln \frac{2q_+ P_-}{Q^2} - \ln P_- R = Y-Y_0 \,.
\end{equation}
We remind the reader that we are using the logarithms of lifetimes${} \cdot P_-$ as rapidities as is conventional. Thus $Y=\ln\frac{2q_+}{Q^2} P_-$ and $Y_0 = \ln R P_-$ with $R$ representing the minimum lifetime, or length, for the nucleus to be effective.

We note that \eqref{eq:3.12} was already found in Eq.~(49) in Ref.~\cite{cite9} with the extra factor of 2 in that reference occurring because both coherent and incoherent processes were not explicitly distinguished. This factor of $Y-Y_0$ in \eqref{eq:3.12} is in many ways surprising, 
just as was the $\frac{1}{\alpha}\ln \frac{Q_s^2}{k_\perp^2}$ term in \eqref{eq:3.5}. In both \eqref{eq:3.5} and \eqref{eq:3.12} one can increase the gluon occupation to arbitrarily large values. In the semiclassical result of \eqref{eq:3.5} the occupation grows as $\ln R$ with $R$ the nuclear size while in \eqref{eq:3.12} one can increase the gluon occupation to arbitrarily large values by making the energy of the reaction arbitrarily large. The $Y-Y_0$ factor in \eqref{eq:3.12} comes from the integration over the $k_+$ variable of the produced gluon shown in Fig.~\ref{fig:3}, or equivalently in Fig.~\ref{fig:7}. This logarithmic integration can be done while fixing the $k_\perp$ and $k_-$ variables, say in Fig.~\ref{fig:7}, which are held fixed in the gluon TMD. As we go to the next level of allowing the nucleus to evolve we shall see that in this more elaborate picture of gluon saturation \eqref{eq:3.5} and \eqref{eq:3.12} are equivalent forms for the gluon TMD and the gluon occupancy.

\subsection{Gluon dipole scattering on an MV nucleus, inelastic scattering} \label{sec:3.3}

\begin{figure}[htbp]
\centering
\begin{tikzpicture}
\begin{feynman} 
\vertex (p1) {$\gamma*$};
\vertex [right=1.5cm of p1] (qa1);
\vertex [above right=0.35cm of qa1] (q1);
\vertex [below right=0.35cm of qa1] (a1);
\vertex [below right=1.2cm of a1] (b1);
\vertex [right=0.2cm of b1] (b2);
\vertex [right=1.5cm of b2] (b3);
\vertex [right=0.75cm of b2] (bm);
\vertex [above=2.0cm of bm] (tt);
\vertex [below=0.8cm of bm] (bb);
\vertex [right=0.2cm of b3] (b4);
\vertex [above right=1.2cm of b4] (a2);
\vertex [above right=0.35cm of a2] (qa2);
\vertex [above left=0.35cm of qa2] (q2);
\vertex [right=1.5cm of qa2] (p2) {$\gamma*$};

\vertex [right=1.6cm of q1] (qm);
\vertex [left=0.75cm of qm] (qml);
\vertex [right=1.6cm of a1] (am);
\vertex [left=0.75cm of am] (aml);

\vertex [above=0.25cm of b2] (n1tl);
\vertex [above=0.25cm of b3] (n1tr);
\vertex [below=0.5cm of n1tl] (n1bl);
\vertex [below=0.5cm of n1tr] (n1br);
\vertex [below left=0.5cm of n1bl] (n1l);
\vertex [below right=0.5cm of n1br] (n1r);
\vertex [right=0.2cm of b2] {$A$};

\diagram*{
  (p1)  -- [charged boson, edge label=$q$] (qa1) -- [quarter left] (q1) -- (qml) -- [fermion, edge label=$\underline{p}$, near end] (qm) -- (q2) -- [quarter left] (qa2) -- [charged boson, edge label=$q$] (p2),
  (qa1) -- [quarter right] (a1) -- (aml) -- [anti fermion, edge label'=$-\underline{p}-\underline{k}$, near start] (am) -- (a2) -- [quarter right] (qa2),
  (a1) -- [charged boson, quarter right, edge label'=$k$] (b1),
  (b4) -- [charged boson, quarter right, edge label'=$k$] (a2),
  (n1tl) -- (n1tr) -- [half left] (n1br) -- (n1bl) -- [half left] (n1tl),
  (n1l) -- [double] (n1bl),
  (n1r) -- [double] (n1br),
};

\diagram* {
  (tt) -- [very thick] (bb),
};
\end{feynman}
\end{tikzpicture}
\caption{ \label{fig:8}}
\end{figure}

We can go from the coherent scattering process where the nucleus does not break up, shown in Figs.~\ref{fig:3} and \ref{fig:7}, to the case of inelastic interactions where the nucleus does break up, by considering the process shown in Fig.~\ref{fig:8} (analogous to Fig.~\ref{fig:3}) and Fig.~\ref{fig:9} (analogous to Fig.~\ref{fig:7}). (We understand that gluon-nucleus scatterings shown in Figs.~\ref{fig:8} and \ref{fig:9} are inelastic so that there is nuclear breaking in both cases.) Now so long as $k_\perp$ is less than the $Q_s$ of the MV nucleus the contributions of the graphs in Fig.~\ref{fig:3} and those of Fig.~\ref{fig:8} are identical. (We draw Fig.~\ref{fig:8} using $x_+$ as a time variable but where an $A_-=0$ gauge is being used so that the gluon dipole cross section, of which Fig.~\ref{fig:3} represents the elastic cross section and Fig.~\ref{fig:8} represents the inelastic gluon dipole-nucleus cross 
section only involve interactions of the nucleus with the soft gluon part of the gluon dipole as is explicit in Fig.~\ref{fig:8}.) The equality of the elastic gluon dipole-nucleus cross section with that of the inelastic dipole-nucleus cross section is completely natural when $k_\perp < Q_s$ and when viewed in the target frame implies the equality of the inelastic and elastic gluon TMD's in the saturation region.

\begin{figure}[htbp]
\centering
\begin{tikzpicture}
\begin{feynman} 
\vertex (tl);
\vertex [below right=1.5cm of tl] (sul);
\vertex [right=0.5cm of sul] (sur);
\vertex [above right=1.5cm of sur] (tr);

\vertex [below=0.0cm of tl] (n1bl);
\vertex [below=0.0cm of tr] (n1br);
\vertex [above=0.5cm of n1bl] (n1tl);
\vertex [above=0.5cm of n1br] (n1tr);
\vertex [below left=0.3535533906cm of n1tl] (n1l);
\vertex [below right=0.3535533906cm of n1tr] (n1r);
\vertex [left=0.5cm of n1l] (n1ll);
\vertex [right=0.5cm of n1r] (n1rr);
\vertex [above right=0.05cm of n1bl] {$A$};

\vertex [below=0.5cm of sul] (sbl);
\vertex [right=0.5cm of sbl] (sbr);
\vertex [below left=1cm of sbl] (p1);
\vertex [below right=1cm of sbr] (p2);
\vertex [right=0.25cm of sul] (sm);
\vertex [above=2.0cm of sm] (tt);
\vertex [above=1.05cm of sm] (tm);
\vertex [below=1.0cm of sm] (bb);

\diagram* {
  (tl) -- [charged boson, edge label'=$k$] (sul) -- (sur) -- [charged boson, edge label'=$k$] (tr),
  (sul) -- (sbl) -- (sbr) -- (sur),
  (p1) -- [charged boson, edge label=$q$] (sbl),
  (sbr) -- [charged boson, edge label=$q$] (p2),
  (n1tl) -- (n1tr) -- [half left] (n1br) -- (n1bl) -- [half left] (n1tl),
  (n1l) -- [double] (n1ll),
  (n1r) -- [double] (n1rr),
};

\diagram* {
  (tt) -- [very thick] (bb),
};
\end{feynman}
\end{tikzpicture}
\caption{ \label{fig:9}}
\end{figure}

\subsection{Gluon dipole scattering on a general nucleus, coherent and inelastic scattering} \label{sec:3.4}

We now repeat the discussion of sections \ref{sec:3.2} and \ref{sec:3.3} but with a general, evolved, nucleus rather than the MV nucleus which was earlier used. In fact little changes in the discussion.

\begin{figure}[htbp]
\centering
\begin{tikzpicture}
\begin{feynman} 
\vertex (p1) {$\gamma*$};
\vertex [right=1.5cm of p1] (qa1);
\vertex [above right=0.35cm of qa1] (q1);
\vertex [below right=0.35cm of qa1] (a1);
\vertex [below right=1.2cm of a1] (b1);
\vertex [right=0.75cm of b1] (b2);
\vertex [right=0.75cm of b2] (bm);
\vertex [right=0.75cm of bm] (b3);
\vertex [above=2.0cm of bm] (tt);
\vertex [below=0.5cm of bm] (bb);
\vertex [right=0.75cm of b3] (b4);
\vertex [above right=1.2cm of b4] (a2);
\vertex [above right=0.35cm of a2] (qa2);
\vertex [above left=0.35cm of qa2] (q2);
\vertex [right=1.5cm of qa2] (p2) {$\gamma*$};

\vertex [right=2.15cm of q1] (qm);
\vertex [left=0.75cm of qm] (qml);
\vertex [right=2.15cm of a1] (am);
\vertex [left=0.75cm of am] (aml);

\vertex [below=1.5cm of b1, blob, minimum height=0.5cm, minimum width=0.5cm] (n1) {$A$};
\vertex [left=0.8cm of n1] (nl1) {$P$};
\vertex [right=1.1cm of n1] (nr1) {$P-\delta$};

\vertex [below=1.5cm of b4, blob, minimum height=0.5cm, minimum width=0.5cm] (n2) {$A$};
\vertex [left=0.5cm of n2] (nl2);
\vertex [right=0.5cm of n2] (nr2);

\diagram*{
  (p1)  -- [charged boson, edge label=$q$] (qa1) -- [quarter left] (q1) -- (qml) -- [fermion, edge label=$\underline{p}$, near end] (qm) -- (q2) -- [quarter left] (qa2) -- [charged boson, edge label=$q$] (p2),
  (qa1) -- [quarter right] (a1) -- (aml) -- [anti fermion, edge label'=$-\underline{p}-\underline{k}$, near start] (am) -- (a2) -- [quarter right] (qa2),
  (a1) -- [charged boson, quarter right, edge label'=$k$] (b1) -- [photon] (bm) -- [charged boson, edge label'=$k+\delta$] (b4) -- [charged boson, quarter right, edge label'=$k$] (a2),
  (b1) -- [gluon] (n1),
  (b4) -- [gluon] (n2),
  (nl1) -- [double] (n1) -- [double] (nr1),
  (nl2) -- [double] (n2) -- [double] (nr2),
};

\diagram* {
  (tt) -- [very thick] (bb),
};
\end{feynman}
\end{tikzpicture}
\caption{ \label{fig:10}}
\end{figure}

In the projectile frame the graph of Fig.~\ref{fig:3} changes to have a more general nucleus as shown 
in Fig.~\ref{fig:10} where we have shown the gluon dipole scattering with the nucleus as it would appear in $A_-=0$ gauge but with $x_+$ being the time variable. The key points are almost identical as in the MV nucleus calculation given in Sec.~\ref{sec:3.2}. With the gluon rapidity $y$ defined by
\begin{equation} \label{eq:3.14}
    y = \ln \frac{2k_+ P_-}{k_\perp^2}
\end{equation}
we require $y<Y = \ln \frac{2 q_+ P_-}{Q^2}$ and we also require $y > \bar{Y}$ where $Q_s^2(\bar{Y}) = k_\perp^2$. That is we restrict the $k_+$-integration only by the lifetime of the $\gamma^*$ and by the requirement that the $q\bar{q}g$ dipole have a strong scattering on the nucleus. Then the only change form the MV nucleus case is that the $Y_0$ in \eqref{eq:3.12} and \eqref{eq:3.13} changes to $\bar{Y}$ and one has the results
\begin{equation} \label{eq:3.15}
    \frac{\dd{N}}{\dd[2]{b}\dd[2]{k}} = \frac{\dd{xG_A}}{\dd[2]{b}\dd[2]{k_\perp}} = \frac{N_c^2-1}{8\pi^4} (Y-\bar{Y}) \,.
\end{equation}
instead of \eqref{eq:3.12} for the coherent gluon TMD of the nucleus. The projectile picture is as given in Fig.~\ref{fig:7}. If the graphs of Fig.~\ref{fig:10} are evaluated in $A_-=0$ gauge, as we have discussed, then the Feynman diagram calculations of the graphs in Fig.~\ref{fig:7} and in Fig.~\ref{fig:10} are identical. The only difference is that in Fig.~\ref{fig:10} we imagine $x_+$ being the time variable while in Fig.~\ref{fig:7} we imagine $x_-$ being the time variable. In a Feynman diagram calculation there is no difference whether we choose $x_+$ or $x_-$ to be the time variable. If we choose $x_-$ to be the time variable then the Feynman graphs are the same as the light cone perturbation theory graphs and so the graphs in Fig.~\ref{fig:7} have a direct partonic interpretation where the gluon $k$ of those graphs is the measured gluon 
giving the coherent TMD. More precisely in Fig.~\ref{fig:7} the $k_\perp$ and $k_-$ of the gluon-$k$ are viewed as measured quantities in the TMD. The variable $k_+$ in Fig.~\ref{fig:7} is in fact negative but it is not a variable of the TMD. In contrast the $k$ in Fig.~\ref{fig:10} has $k_+ > 0$ and the integration of the $k_+$ give the $Y-\bar{Y}$ factor while in Fig.~\ref{fig:7} the $Y-\bar{Y}$ factor is, in a light cone perturbation theory calculation, from the $\delta_-$-integration of the $\delta-k$ line.

As in the case of the MV nucleus the inelastic gluon TMD is the same as the coherent TMD and so is also given by \eqref{eq:3.15}.

Thus, here again, we have the, perhaps, surprising result that by making the energy of the $\gamma^*$-nucleus scattering very large we can make the coherent and inelastic gluon TMD's as large 
as we wish. There is no $1/\alpha$ factor appearing directly in \eqref{eq:3.15} however we can change the form of the TMD's, or gluon occupations, by writing $Y-\bar{Y}$ in \eqref{eq:3.15} in terms of the nucleon's saturation momentum. We do that here in a fixed coupling calculation of $Q_s$. (Somewhat later we will introduce a running coupling calculation of $Q_s$.) Then writing \cite{cite1,cite28,cite29}
\begin{equation} \label{eq:3.16}
    Q_s^2(Y) = Q_s^2(\bar{Y}) e^{\frac{2\alpha N_c}{\pi} \frac{\chi(\lambda_0)}{1-\lambda_0} (Y-\bar{Y})}
\end{equation}
the coherent and inelastic gluon TMD's of the nucleus can be written as
\begin{equation} \label{eq:3.17}
    \frac{\dd{xG_A}}{\dd[2]{b}\dd[2]{k_\perp}} = \frac{N_c^2-1}{8\pi^4} \qty[\frac{2\alpha N_c}{\pi} \frac{\chi(\lambda_0)}{1-\lambda_0}]^{-1} \ln \frac{Q_s^2(Y)}{k_\perp^2}
\end{equation}
where we have $Q_s^2(\bar{Y}) = k_\perp^2$ as discussed above. In \eqref{eq:3.16} and \eqref{eq:3.17}, $\chi$ is the usual BFKL \cite{cite30Balitsky,cite30Fadin,cite31} characteristic function with $\lambda_0$ defined by  
\begin{equation} \label{eq:3.18}
    \frac{\chi(\lambda_0)}{\chi'(\lambda_0)} = -(1-\lambda_0) \,.
\end{equation}

Except for constant factors \eqref{eq:3.17} is just like \eqref{eq:3.5} although the physics giving \eqref{eq:3.5} and \eqref{eq:3.17} are very different. Eq.~\eqref{eq:3.5} comes about completely from multiple scattering leading to a transverse broadening spectrum depending on where in the nucleus the current hits a nucleon (projectile picture). There are no contributions coming from integrations over longitudinal momentum. On the other hand the whole contribution of \eqref{eq:3.17} comes from the various longitudinal momentum values $k_+$, both in a projectile and in a target picture. For a given $k_+$ unitarity dictates the contribution to $\frac{\dd{N}}{\dd[2]{b}\dd[2]{k_\perp}}$, and the large value for that TMD comes from all the different $k_+$'s for which unitarity for the gluon dipole is reached. Since elastic and inelastic dipole nucleus scattering 
are the same when the unitarity limit is reached the coherent (elastic) and inelastic TMD's are exactly the same values when $k_\perp / Q_s < 1$.

\subsection{Possible additional contributions to gluon saturation and to the gluon TMD in the saturation region} \label{sec:3.5}

In the next section we shall discuss in detail Sudakov effects and how they ultimately limit gluon occupancies, but at a surprisingly large value. In this section we are going to look at possible additions to the contributions we have already considered to see that they all (except Sudakov effects) appear to be small.

We have taken the $q\bar{q}$ pair to have a large relative transverse momentum and this is our choice in order to have a good measurer of the gluon as seen, for example, in Fig.~\ref{fig:7} for the coherent gluon TMD. So far we have considered only one 
gluon $k$, with $k_\perp < Q_s$, in Fig.\ref{fig:7} and this is the measured gluon in the TMD. What about having additional gluons? In particular suppose in addition to gluon $k$ there is a second gluon $l$, having $l_\perp > k_\perp$, being emitted from the $q\bar{q}$ pair (in the projectile frame) before the gluon ($k$) is emitted. In this case we would be considering a gluon TMD of transverse momentum $l_\perp$ rather than a TMD of transverse momentum $k_\perp$ so we reject this possibility. (A gluon $l$ having $Q>l_\perp>k_\perp$ and emitted after the gluon $k_\perp$, in the projectile frame, will have a real-virtual cancellation.)

Suppose, on the other hand there is an additional soft gluon, $l$, having $l_+ < k_+$ and $\tau_l < \tau_k$, so that the scattering with the target can be either that of a $q\bar{q}g(k)g(l)$ system or, for final state $l$ emissions, that of a $q\bar{q}g(k)$ system. 
As shown in detail in Refs.~\cite{cite31,cite32}  there is an initial state-final state emission cancellation of the gluon $l$ which occurs leaving us, again, with only a $q\bar{q}g(k)$ system scattering on the target. And this is exactly the circumstance we have been considering in the previous sections of this paper.

Our conclusion here that the only scattering relevant to get the gluon nuclear TMD for a gluon having $(k_-,k_\perp)$ is that of the scattering of the $q\bar{q}g(k)$ system on the target. This can also be argued by the following: In the projectile frame the $\gamma^*$ first breaks up into a compact $q\bar{q}$ system. However this system is very weakly interacting so an additional gluon must correspondingly be sensitive to gluon saturation. Thus sometime before reaching the target the 
$q\bar{q}$ emits a gluon of transverse momentum $k_\perp<Q_s$ and longitudinal momentum $k_+$. This $q\bar{q}g(k)$ system is now strongly interacting and we may consider its further evolution before reaching the target that of the evolution of the $\ket{q\bar{q}k(k)+}$ scattering state. Since the $q\bar{q}g(k)$ system is strongly interacting we expect two classes of events in the scattering of $\ket{q\bar{q}g(k)+}$ on the target \cite{cite33,cite34}: (i) There will be highly inelastic events where the nucleus breaks up. (ii) There will be coherent events where the nucleus remains in its ground state and a system of (three) $q$, $\bar{q}$ and $g$ jets is diffractively produced. At the time of scattering, and in the projectile frame, there may be other gluons (besides $g(k)$) moving in the projectile direction but the additional gluons will not be produced since they are not part of 
the initial $q\bar{q}g$ state coming from the $\gamma^*$. Thus graphs having additional real gluons beyond that shown in Fig.~\ref{fig:10} will cancel among themselves. And that means corrections beyond what we have already calculated must be virtual graphs and it is to these contributions that we now turn.

\section{Sudakov effects} \label{sec:4}
\subsection{Fixed coupling Sudakov effects} \label{sec:4.1}

As we have discussed in some detail the large value of the gluon TMD, say the coherent TMD, given in \eqref{eq:3.15} and \eqref{eq:3.17} comes from the equivalent sets of graphs shown in Fig.~\ref{fig:7} and Fig.~\ref{fig:10}. Let's focus on the graphs in Fig.~\ref{fig:10}. The fact that they represent a gluon TMD of transverse momentum $k_\perp$ is evident in the equivalent graphs of Fig.~\ref{fig:7} where the $q\bar{q}$ pair attached to the $\gamma^*$ serves as a measurer of the gluon and of its $k_\perp$ and 
$k_-$-values. (The $k_-$ is always fixed by the requirement that the $q\bar{q}$ pair be on shell.) The large contribution to \eqref{eq:3.15} comes directly from the $\frac{\dd{k_+}}{k_+}$ integration giving the $(Y-\bar{Y})$ factor in \eqref{eq:3.15}. When $Y-\bar{Y}$ is large it means that there is typically a very large rapidity gap between the $q\bar{q}$ pair and the gluon $g(k)$. The emission of a gluon $l$ (with $l_\perp > k_\perp$ and $\frac{2l_+}{l_\perp^2} > \frac{2k_+}{k_\perp^2}$) in addition to the gluon $k$, would partially fill in the gap between $q\bar{q}$ and $g(k)$ but, as we argued earlier, this would correspond to a TMD for gluon $l$ instead of gluon $k$ so we dismissed such an emission. Thus the requirement that there be no gluon emission, $l$, in the time region $\frac{2k_+}{k_\perp^2} < \frac{2l_+}{l_\perp^2} < \frac{2q_+}{Q^2}$ has a small probability which will diminish the result \eqref{eq:3.15}. In fact this diminishing of \eqref{eq:3.15} due to the lack of the gluon emission $l$ is taken into account by virtual corrections to the graphs in Fig.~\ref{fig:7} and Fig.~\ref{fig:10}. 
The virtual corrections which diminish the TMD are shown in Fig.~\ref{fig:11}. These contributions are of the Sudakov type \cite{cite18,cite19,cite20,cite21,cite22}. The $l$-loop contribution shown in Fig.~\ref{fig:11}, along with an identical contribution in the complex conjugate amplitude, give the Sudakov factor at lowest order,

\begin{figure}[htbp]
\centering
\begin{tikzpicture}
\begin{feynman}
\vertex [blob, minimum height=0.5cm, minimum width=0.5cm] (n1) {};
\vertex [left=0.8cm of n1] (nl1) {$P$};
\vertex [right=1.1cm of n1] (nr1) {$P-\delta$};
\vertex [below=1.5cm of n1] (tl);
\vertex [right=2.5cm of tl] (tm);
\vertex [below right=1.0cm of tl] (l1);
\vertex [below right=1.5cm of l1] (l2);
\vertex [below right=0.5cm of l2] (sul);
\vertex [right=0.5cm of sul] (sur);
\vertex [above right=3.0cm of sur] (tr);
\vertex [right=4.72cm of n1, blob, minimum height=0.5cm, minimum width=0.5cm] (n2) {};
\vertex [left=1.1cm of n2] (nl2) {$P-\delta$};
\vertex [right=0.8cm of n2] (nr2) {$P$};
\vertex [below=0.5cm of sul] (sbl);
\vertex [right=0.5cm of sbl] (sbr);
\vertex [below left=1cm of sbl] (p1);
\vertex [below right=1cm of sbr] (p2);
\vertex [right=0.25cm of sul] (sm);
\vertex [above=2.6cm of sm] (tt);
\vertex [below=1.0cm of sm] (bb);

\diagram* {
  (nl1) -- [double] (n1) -- [double] (nr1),
  (nl2) -- [double] (n2) -- [double] (nr2),
  (n1) -- [gluon] (tl) -- [photon] (tm) -- [charged boson, edge label=$\delta-k$] (tr) -- [gluon] (n2),
  (tl) -- [charged boson, edge label'=$k$] (l1) -- [charged boson, edge label=$k-l$] (l2) -- [photon] (sul) -- (sur) -- [photon] (tr),
  (sul) -- (sbl) -- (sbr) -- (sur),
  (l1) -- [charged boson, half right, edge label'=$l$] (l2),
  (p1) -- [charged boson, edge label=$q$] (sbl),
  (sbr) -- [charged boson, edge label=$q$] (p2),
};

\diagram* {
  (tt) -- [very thick] (bb),
};
\end{feynman}
\end{tikzpicture}
\caption{ \label{fig:11}}
\end{figure}

\begin{equation} \label{eq:4.1}
    \mathit{Sud}^{(1)} = -\frac{2\pi g^2N_c}{(2\pi)^3} \int_{k_\perp^2}^{Q^2} \frac{\dd{l_\perp^2}}{l_\perp^2} \int_{l_\perp^2/q_+}^{k_-} \frac{\dd{l_-}}{l_-} \,.
\end{equation}
Thus
\begin{equation} \label{eq:4.2}
    \mathit{Sud}^{(1)} = -\frac{\alpha N_c}{2\pi} \ln^2\frac{Q^2}{k_\perp^2} \,.
\end{equation}
The limits of integration in \eqref{eq:4.1} come from the following: (i) The upper limit, $Q^2$, of the $l_\perp^2$-integration comes from renormalization. (ii) The lower limit $k_\perp^2$ of the $l_\perp^2$-integration is because $l_\perp < k_\perp$ contributions are canceled by corresponding real emissions. (iii) The limit $l_- < k_-$ is obvious. (iv) The limit $l_- > l_\perp^2/q_+$ is the condition that the lifetime of the $l$-line be longer 
than that of the quark loop. The correction \eqref{eq:4.2} naturally exponentiates to give the full Sudakov factor
\begin{equation} \label{eq:4.3}
    \mathit{Sud} =  e^{-\frac{\alpha N_c}{2\pi} \ln^2\frac{Q^2}{k_\perp^2}} \,.
\end{equation}
Then adding the Sudakov factor to our previous fixed coupling result gives our final answer, in the fixed coupling approximation, for the coherent gluon TMD of a hadron or nucleus as \cite{cite16}
\begin{equation} \label{eq:4.4}
    \frac{\dd{N}}{\dd[2]{b} \dd[2]{k_\perp}} = \frac{\dd{xG_A}}{\dd[2]{b}\dd[2]{k_\perp}} = \frac{N_c^2-1}{(2\pi)^4}\ \frac{1-\lambda_0}{\chi(\lambda_0)} \frac{1}{\alpha N_c/\pi} \ln\frac{Q_s^2}{k_\perp^2} e^{-\frac{\alpha N_c}{2\pi} \ln^2\frac{Q^2}{k_\perp^2}} \,.
\end{equation}

We note the strong dependence of the Sudakov factor, and with it the coherent gluon TMD, on the $Q^2$ of the process. If one were calculating the full coherent gluon distribution of, say, a nucleus then real gluons could be emitted in the graph of Fig.~\ref{fig:11} with $k_\perp^2<l_\perp^2<Q^2$ exactly canceling 
the Sudakov factor in \eqref{eq:4.1}. However, in the TMD these real gluons are not allowed, and the larger $Q^2$ becomes the more real gluon emissions are being suppressed. The value of $Q^2$ is our choice. So long as $Q^2/Q_s^2 \gg 1$ the $\gamma^*$ along with its accompanying $q\bar{q}$ system is an effective measurer of gluons having $k_\perp$ in a nucleus with saturation momentum $Q_s$. Thus, parametrically, one can take $Q^2$ as small as $N_0 Q_s^2(Y)$ and still have an effective measurement of the coherent gluon TMD. $N_0$ is just some large, $N_0 \gg 1$, pure number which we shall drop in the formulas which follow because the $N_0$ corrections are subleading.

Now let's go back to \eqref{eq:4.4} and ask where are the gluon occupancies the largest and how large they can be. Start at $k_\perp$ just less than $Q_s$ where $k_\perp/Q_s$ is on the order of one. Here occupancies are on the order of $1/\alpha$. The 
occupancy increases as one lowers $k_\perp^2$ giving the formula \eqref{eq:3.17} until the exponential in \eqref{eq:4.4} becomes large. The peak occupancy occurs when 
\begin{equation} \label{eq:4.5}
    \dv{\ln k_\perp^2} \qty(\frac{\dd{N}^{coh}}{\dd[2]{b}\dd[2]{k_\perp}}) = 0
\end{equation}
which occurs when
\begin{equation} \label{eq:4.6}
    \frac{\alpha N_c}{\pi} \ln^2 \frac{Q_s^2}{k_\perp^2} = 1
\end{equation}
or
\begin{equation} \label{eq:4.7}
    \ln \frac{Q_s^2}{k_\perp^2} = \sqrt{\frac{\pi}{\alpha N_c}}
\end{equation} \,.
This gives a maximum value for the coherent occupancy as
\begin{equation} \label{eq:4.8}
    \eval{\frac{\dd{N}^{coh}}{\dd[2]{b}\dd[2]{k_\perp}}}_{max} \simeq \frac{N_c^2-1}{(2\pi)^4}\ \frac{1-\lambda_0}{\chi(\lambda_0)} \qty[\frac{\pi}{\alpha N_c}]^{3/2} \frac{1}{\sqrt{e}} \,.
\end{equation}
It is straightforward to check that the same value, \eqref{eq:4.8}, also holds for the maximum value of the occupancy for inelastic reactions \cite{cite16}.

Now, 
\begin{equation} \label{eq:4.9}
    \frac{\dd{N}}{\dd[2]{b}\dd[2]{k_\perp}} \sim \frac{1}{\alpha^{3/2}}
\end{equation}
is a surprisingly large number for gluon occupancies in the light cone wavefunction of a hadron or nucleus. One might have thought that when the occupancy becomes as large as $1/\alpha$ that the rate of production of saturated gluons would cease and that gluon recombinations (2 gluons $\to$ 1 gluon) would limit the occupancy to values on the order of $1/\alpha$. However, as we have seen the driving force for large occupancies is the $\frac{\dd{k_+}}{k_+}$ integration seen in Fig.~\ref{fig:7} and especially manifest in the graphs of Fig.~\ref{fig:10}. This same gluon of the coherent TMD whose values of $k_\perp$ and $k_-$ are fixed by the TMD can have a wide range of $k_+$ values, the range of $\ln k_+$ growing linearly with the rapidity, $Y$, of the process thus is driving the growth of the TMD.

Thus, as far as we understand the dominant contribution to the coherent TMD comes from graphs of the type shown in Fig.~\ref{fig:7} and Sudakov limited by graphs of the type in Fig.~\ref{fig:11}. In each case there is a single ``struck gluon'' produced in addition to the $q\bar{q}$ pair of jets and the nucleus. Of course at the time the $q\bar{q}$ pair absorbs the gluon, gluon $k$ in Fig.~\ref{fig:7}, there may be many additional gluons present but they are part of the nuclear wavefunction and do not appear as final state jets. We shall come back to this a little later.

\subsection{Running coupling Sudakov effects} \label{sec:4.2}
We now redo the discussion of Sudakov effects but now with a running QCD coupling rather than fixed coupling dynamics. As we shall see the Sudakov effects for the gluon TMD are essentially the same using a running coupling 
as well as for fixed coupling dynamics \cite{cite16}.

We begin with the lowest order Sudakov contribution given by \eqref{eq:4.1} and \eqref{eq:4.2} in the fixed coupling limit. The running coupling is taken into account in \eqref{eq:4.1} by replacing $\alpha$ in that equation by
\begin{equation} \label{eq:4.10}
    \alpha(l_\perp^2) = \frac{1}{b\ln(l_\perp^2/\Lambda^2)}
\end{equation}
where $b = (11 N_c - 2N_f) / 12\pi$ so that \eqref{eq:4.1} becomes
\begin{equation} \label{eq:4.11}
    \mathit{Sud}^{(1)} = -\frac{N_c}{\pi b} \int_{k_\perp^2}^{Q^2} \frac{\dd{l_\perp^2}}{l_\perp^2} \frac{1}{\ln(l_\perp^2/\Lambda)} \int_{l_\perp^2/q_+}^{k_-} \frac{\dd{l_-}}{l_-}
\end{equation}
which gives
\begin{equation} \label{eq:4.12}
    \mathit{Sud}^{(1)} = -\frac{N_c}{\pi b} \qty[\ln \frac{Q^2}{\Lambda^2} \ln(\frac{\ln (Q^2/\Lambda^2)}{\ln (k_\perp^2/\Lambda^2)}) - \ln \frac{Q^2}{k_\perp^2}]
\end{equation}
which can be better written as
\begin{equation} \label{eq:4.13}
    \mathit{Sud}^{(1)} = -\frac{N_c}{\pi b} \qty[\ln\frac{Q^2}{\Lambda^2} \ln(1+\frac{\ln(Q^2/k_\perp^2)}{\ln(k_\perp^2/\Lambda^2)}) - \ln\frac{Q^2}{k_\perp^2}]
\end{equation}
with the full Sudakov factor being the exponential of \eqref{eq:4.13}. As in the fixed coupling case we take $Q^2/Q_s(Y)^2$ somewhat greater than one so that gluons of interest are 
well measured by our probe at scale $Q^2$. Also, in order to avoid a strong Sudakov suppression we take $\frac{\ln(Q^2/k_\perp^2)}{\ln(k_\perp^2/\Lambda^2)} \ll 1$ in \eqref{eq:4.13} so that
\begin{equation} \label{eq:4.14}
    \mathit{Sud}^{(1)} \simeq -\frac{N_c}{2\pi b} \frac{\ln^2(Q^2/k_\perp^2)}{\ln(k_\perp^2/\Lambda^2)} \simeq -\frac{N_c}{2\pi b} \frac{\ln^2(Q_s^2(Y)/k_\perp^2)}{\ln(k_\perp^2/\Lambda^2)}
\end{equation}
which gives the full Sudakov factor as
\begin{equation} \label{eq:4.15}
    \mathit{Sud} = e^{-\frac{N_c}{2\pi b} \frac{\ln^2(Q_s^2(Y)/k_\perp^2)}{\ln(k_\perp^2/\Lambda^2)}} \,.
\end{equation}

Taking the running coupling Sudakov factor in \eqref{eq:4.15} times \eqref{eq:3.15} and replacing $\ln(k_\perp^2/\Lambda^2)$ by $\ln(Q_s^2/\Lambda^2)$ gives
\begin{equation} \label{eq:4.16}
    \frac{\dd{N}}{\dd[2]{b} \dd[2]{k_\perp}} = \frac{N_c^2-1}{8\pi^4} (Y-\bar{Y}) e^{-\frac{N_c}{2\pi b} \frac{\ln[2](Q_s^2(Y)/k_\perp^2)}{\ln(Q_s^2(Y)^2/\Lambda^2)}} \,.
\end{equation}
Finally, we should use the running coupling result also for the saturation momentum \cite{cite1,cite29,cite35}
\begin{equation} \label{eq:4.17}
    \ln\frac{Q_s^2(Y)}{\Lambda^2} \simeq \sqrt{\frac{4N_c}{\pi b} \frac{\chi(\lambda_0)}{1-\lambda_0} Y} \,,
\end{equation}
and, using \eqref{eq:4.17} to evaluate $Y-\bar{Y}$ in \eqref{eq:4.16} in terms of the saturation momentua one finds 
\begin{equation*}
    Y-\bar{Y} = \qty[\frac{4N_c}{\pi b} \frac{\chi(\lambda_0)}{1-\lambda_0}]^{-1} \qty(\ln^2\frac{Q_s^2(Y)}{\Lambda^2} - \ln^2\frac{k_\perp^2}{\Lambda^2})
\end{equation*}
or
\begin{equation} \label{eq:4.18}
    Y-\bar{Y} = \qty[\frac{4N_c}{\pi b} \frac{\chi(\lambda_0)}{1-\lambda_0}]^{-1} 2\ln\frac{Q_s^2(Y)}{k_\perp^2} \ln\frac{Q_s^2(Y)}{\Lambda^2}
\end{equation}
Then, using \eqref{eq:4.18} in \eqref{eq:4.16} one finds
\begin{equation} \label{eq:4.19}
    \frac{\dd{N}}{\dd[2]{b} \dd[2]{k_\perp}} = \qty[\frac{2N_c}{\pi b} \frac{\chi(\lambda_0)}{1-\lambda_0}]^{-1} \frac{N_c^2-1}{8\pi^4} \ln\frac{Q_s^2(Y)}{\Lambda^2} \ln\frac{Q_s^2(Y)}{k_\perp^2} e^{-\frac{N_c}{2\pi b} \frac{\ln[2](Q_s^2(Y)/k_\perp^2)}{\ln(Q_s^2(Y)/\Lambda^2)}} \,.
\end{equation}
Using the formula, \eqref{eq:4.10}, for the running coupling one finally gets
\begin{equation} \label{eq:4.20}
    \frac{\dd{N}}{\dd[2]{b} \dd[2]{k_\perp}} = \frac{N_c^2-1}{8\pi^4} \frac{1-\lambda_0}{2\chi(\lambda_0)} \frac{1}{\alpha N_c/\pi} \ln\frac{Q_s^2(Y)}{k_\perp^2} e^{-\frac{\alpha N_c}{2\pi} \ln^2\frac{Q_s^2(Y)}{k_\perp^2}} \,.
\end{equation}
We note that \eqref{eq:4.20} is identical to \eqref{eq:4.4} with the $\alpha$'s in \eqref{eq:4.20} understood to be $\alpha(Q_s^2(Y))$ while the $\alpha$'s in \eqref{eq:4.4} are fixed coupling. This means that again the occupations can become as large as $\alpha^{-3/2}(Q_s^2(Y))$.

Thus, there appears to be no essential difference between fixed coupling and running coupling 
as far as the properties of saturation and the occupancy of gluons in the light cone wavefunction of a high energy nucleus (or nucleon).

\section{Key issues in gluon saturation} \label{sec:5}

Quark and gluon saturation are rather highly developed and mature subjects. Nevertheless there are still many issues which are only partially understood. There are also issues where one seems to know technical results but where the underlying physical picture is still mysterious or confusing, while there are other issues where the physical picture is becoming clearer but where that picture has not been widely discussed in the literature. It is to some of these diverse topics that we now turn.

\subsection{Coherent states} \label{sec:5.1}

Coherent quark-antiquark states were discussed 
in some detail in Ref.~\cite{cite13}. Here we focus on coherent gluon pairs in the light cone wavefunction, say, of a large nucleus. We first remind the reader of the idea of a coherent state in the simple case where $a^\dagger$ creates a particle and $a$ destroys a particle which has no momentum or other quantum numbers. Then
\begin{equation} \label{eq:5.1}
    [a,a^\dagger] = 1 \,.
\end{equation}
A coherent state is formed as
\begin{equation} \label{eq:5.2}
    \ket{\psi_\lambda} = e^{\lambda a^\dagger} \ket{0}
\end{equation}
where $a\ket{0} = 0$. It is easy to check that 
\begin{equation} \label{eq:5.3}
    a\ket{\psi_\lambda} = \lambda \ket{\psi_\lambda} \,.
\end{equation}
Then $\ket{\psi_\lambda}$ is the eigenstate of an annihilation operator and thus a coherent state. Now let's see how this idea applies to coherent diffractive scattering in the saturation 
region in high energy QCD.

We consider the scattering of a $\ket{q\bar{q}g(k)+}$ state, coming from a $\gamma^*$, on a large nucleus as discussed in great detail in Secs.~\ref{sec:3.4} and \ref{sec:3.5}. As before we imagine the transverse momentum, $p_\perp$, of the incident quark to have $p_\perp/Q_s \gg 1$ while the transverse momentum of the gluon $k_\perp$ is less than $Q_s$. (The transverse momentum of the antiquark, $\bar{q}$, is then close to $-p_\perp$ and the scattering on the nucleus in the projectile frame looks like that of a gluon dipole scattering on the nucleus \cite{cite11,cite12,cite14}.) Using $A_-=0$ gauge, but where $x_+$ is the time variable, the process is as illustrated in Fig.~\ref{fig:10} where in $A_-=0$ gauge the soft gluon part of the ``gluon'' dipole does the scattering. In the target frame, where $x_-$ 
is the time variable, the graphs of Fig.~\ref{fig:10} change into those of Fig.~\ref{fig:7}. (The graphs are the same but they change appearance because of the change in the time variable.)

When $k_\perp < Q_s$ the coherent reaction of Fig.~\ref{fig:10} is at the unitarity limit for the gluon dipole for any value of $k_+$ which keeps $\ln \tau_k P_- > \bar{Y}$ with $Q_s(\bar{Y}) = k_\perp$. After integrating over $k_+$, but still for fixed $k_\perp$ the result \eqref{eq:3.17}, illustrated in Fig.~\ref{fig:7}, represents the coherent gluon TMD at $k_- = -q_-$ and $k_\perp$. There is a peculiarity in the graphs of Fig.~\ref{fig:7} when looked at closely. The event happens when the gluon $k$ is absorbed by the $q\bar{q}$ system connected to the $\gamma^*$. The $q\bar{q}g$ and $q\bar{q}\gamma$ vertices happen at almost the same values of $x_-$ and so the $q\bar{q}\gamma$ which is the measurer of the gluon $k$ acts essentially instantaneously in the absorption of $k$. Just before 
$k$ is absorbed the light cone wavefunction is that of the nucleus of momentum $P$. Shortly after the gluon $k$ is absorbed the system consists of the nucleus $A$, the gluon $(\delta-k)$ and the $q\bar{q}$ pair of jets. Over the short period of time that the gluon $k$ disappears the nuclear wavefunction does not change so that we may also say that just before the gluon $k$ is absorbed the light cone wavefunction is the light cone wavefunction of $A$ along with the gluons $(\delta-k)$ and $(k)$. In a coherent diffractive reaction the measurement of gluon $k$ is equivalent to measuring the two gluon system $k$ and $\delta-k$ (but not the $(-)$ component of the second gluon, $\delta-k$) so that measuring the $g(k)$, $g(\delta-k)$ part of the nuclear wavefunction leaves the nuclear wavefunction unchanged exactly as expressed in \eqref{eq:5.3}, the criterion for a coherent state.

From the projectile point of view, Fig.~\ref{fig:10}, the fact that the initial nucleus remains unchanged 
is just unitarity, in the strong scattering regime, for the state $\ket{q\bar{q}g+}$ on the nucleus. This ``elastic'' scattering of the $q\bar{q}g$ system on the nucleus is something we are familiar with and expect. The target picture of Fig.~\ref{fig:7} is less familiar, but seems well covered by saying that the nucleus is a coherent state for pairs of gluons, $\delta-k$ and $k$, having $k_\perp < Q_s$ and $\delta_-$ variable with only the requirement that $\delta_-$ not be so large as to bring $k$ out of the saturation region for the scattering shown in Fig.~\ref{fig:10}.

While the idea of the existence of a coherent state, a target frame concept (see Fig.~\ref{fig:7}) si fairly clear its existence and construction are done following projectile frame calculations (see Fig.~\ref{fig:10}). What is missing here is a fully target frame understanding of why coherent gluon states must exist and how to evaluate 
them.

\subsection{Shadowing} \label{sec:5.2}

The issue here is the relationships between the shadowing of the gluon TMD and the gluon occupancies given by the TMD. We shall again focus on coherent TMD's although all we conclude for coherent TMD's is also true for the inelastic gluon TMD. However, we begin with the example from Sec.~\ref{sec:3.2}, the quasiclassical gluon, where there are only inelastic reactions with no coherent reactions as is manifest from Fig.~\ref{fig:4}.

\subsubsection{Quasiclassical gluon} \label{sec:5.2.1}

Fig.~\ref{fig:4} also makes clear that the gluon distribution will have no shadowing. The initial reaction, in the projectile frame, where the ``current'' produces gluon $(q+\delta_1)$ can happen any place within the nucleus. After being 
produced the gluon $(q+\delta_1)$ can then carry out multiple scattering but it never goes away. The surprise then is that this reaction, where the integrated gluon distribution exhibits no shadowing as explicitly shown in \eqref{eq:3.11}, nevertheless has an inelastic TMD, \eqref{eq:3.2} and \eqref{eq:3.3}, which except for an overall factor is identical to the TMD given in \eqref{eq:3.17} where that form is intimately related to the unitarity scattering of the gluon dipole exhibited in Fig.~\ref{fig:3} and in Fig.~\ref{fig:10}. The reason for this close relationship between the TMD for the quasiclassical gluon where the transverse momentum distribution is given by random multiple scattering (see Fig.~\ref{fig:4}) and the full QCD answer (before Sudakov effects) is that the
random multiple scattering is given by \eqref{eq:3.3} with the $(1-e^{-\underline{x}^2Q_s^2/4})$ in that formula being the same as coming from the $T$-matrix for the scattering of a gluon dipole of size $x_\perp$ on a nucleus with saturation momentum $Q_s$ \cite{cite23}.

The fact that the quasiclassical gluon and a full quantum gluon have identical \textit{forms} in the $k_\perp$-dependence of their TMD's while the full quantum gluon has shadowing while the quasiclassical gluon has no overall shadowing is possible because of the very different formulas for the saturation momentum. For the quasiclassical gluon the saturation momentum is given by \eqref{eq:3.4} while for the full quantum gluon, but
for fixed coupling, the saturation momentum is given by \eqref{eq:3.16}. Thus the forms of \eqref{eq:3.17} and \eqref{eq:3.5} are the same but the saturation momenta in these two formulas are very different. In each case the gluon TMD is strongly modified in going from a nucleon to a nucleus. The functional forms of the TMD's for the semiclassical and full quantum gluons are identical. However when the $k_\perp$ is integrated over to get the gluon distributions the semiclassical case has no shadowing while there is strong shadowing in the case of the full quantum gluon.

\subsubsection{Quantum gluons} \label{sec:5.2.2}

We now turn to quantum gluons, described in Sec.~\ref{sec:3.2} and \ref{sec:3.4}. We always suppose the gluon of the TMD that concerns us is in the saturation regime in which case there is not much difference between a MV nucleus (Sec.~\ref{sec:3.2}) and a fully evolved nucleus (Sec.~\ref{sec:3.4}). Nevertheless, the shadowing we are describing in a MV nucleus is purely leading twist shadowing. If the gluon whose shadowing we are describing is viewed in the projectile frame then that gluon is the soft gluon of the gluon dipole-nucleus scattering. Shadowing then comes completely from the gluon dipole scattering on a set of essentially free nucleons. But in $A_-=0$ gauge the interaction 
of the gluon dipole with the target comes about completely through interactions of the soft gluon with the target as shown in Fig.~\ref{fig:10}. Thus there are only two gluons interacting with the quark loop which means the process is leading twist. Remaining in $A_-=0$ gauge but with $x_-$ as the time variable the graphs of Fig.~\ref{fig:10} become the graphs of Fig.~\ref{fig:7} where the leading twist nature of the shadowing is manifest.

The process is leading twist but shadowing comes about because the gluon dipole scattering on the nucleus, shown in Fig.~\ref{fig:10}, is limited by unitarity to be much less than the sum of the dipole-nucleon cross sections where all the nucleons at the same impact parameter of the dipole are included.

Now at first glance one might suppose that the gluon shadowing which comes from the unitarity limits imposed on the gluon dipole-nucleus 
scattering would also lead to limits on the size of the gluon TMD or, equivalently, on the gluon occupancy in the light cone wavefunction of the nucleus. Indeed, this is exactly what happens for the quark TMD given earlier in \eqref{eq:2.14} and illustrated in Fig.~\ref{fig:2}. The strong shadowing (unitarity) which limits the quark-antiquark dipole-nucleus scattering also limits the quark occupancy in the light cone wavefunction of the nucleus to be on the order of one. However, this is not the case for the gluon occupancy.

Consider the projectile frame picture, but in $A_-=0$ light cone gauge as shown in Fig.~\ref{fig:10}. This gluon dipole-nucleus scattering is exactly analogous to the quark dipole-nucleus scattering shown in Fig.~\ref{fig:2}. In each case the multiple scattering of the softer part of the dipole on the nucleus 
limits the gluon, or quark, occupancy in the corresponding nuclear wavefunction. In the quark dipole case fixing $p_-$ and $p_\perp$, which are the variables of the TMD, effectively fixes $p_+$ also. However in the gluon dipole case fixing $k_-$ and $k_\perp$, as shown in Fig.~\ref{fig:10} or Fig.~\ref{fig:7} does not fix $k_+$. The unitarity restriction limits the occupancy for fixed $k_+$, in addition to fixed $k_-$, $k_\perp$, to be on the order of one, but the integration over $k_+$ gives the $(Y-\bar{Y})$ factor in \eqref{eq:3.15} and thus allows gluon occupancies to become arbitrarily large, before Sudakov effects. Thus the gluon occupancy is lowered but not limited by unitarity in the projectile frame. We know of no mechanism, besides Sudakov factors, which limit gluon occupancies. In particular shadowing,
equivalent to unitarity, can lower but not limit occupancies in the nuclear wavefunction.

\subsection{Initial condition for a high energy heavy ion collision} \label{sec:5.3}

We recall the initial condition for a high energy heavy ion collision in the central region of the collision. In the center of mass frame for the collision and for the low longitudinal momentum components of the nuclear wavefunctions which have high occupancy in the saturation regions of the two colliding nuclei we expect strong interactions to occur. There strong interactions are expected to free all these soft gluons and these soft gluons become the initial state of the system which evolves into the quark-gluon plasma. In Ref.~\cite{cite24} the evolution of this system was parametrically, but analytically, evaluated.
Various stages in the evolution of the initial, high occupancy, system of gluons up to equilibration were identified. Later numerical studies \cite{cite26} confirmed this evolution toward equilibrium of the quark-gluon plasma. In these analytic estimates one assumed that the initial condition consisted of gluons having $k_\perp\sim Q_s$ and occupation number of size $1/\alpha$. We have seen now that occupations as large as $1/\alpha^{3/2}$ should be present in the initial condition. Do we expect this to change the earlier conclusions as to the early stages of evolution after a heavy ion collision?

In fact the very high occupancy gluons are not very important in a heavy ion collision. Let's do a rough estimate to see that this is the case. At the time of the collision the
transverse energy carried by gluons having transverse momentum $k_\perp$ is
\begin{equation} \label{eq:5.4}
    E_T = -\pi R^2 \cdot \frac{1}{\pi/k_\perp^2} \cdot k_\perp \cdot f(k_\perp) = 4R^2k_\perp^3 f(k_\perp)
\end{equation}
where in our simple estimate the first factor, $\pi R^2$, is the area of the nucleus, $\pi/k_\perp^2$ is the area of a gluon of transverse momentum $k_\perp$, the third factor $k_\perp$ is the transverse energy of a gluon while the final factor, $f(k_\perp)$, is the occupancy of gluons having $k_\perp$ at the rapidity in question. Referring to \eqref{eq:4.4} we see that the growth of $f =  \frac{\dd{N}}{\dd[2]{b} \dd[2]{k_\perp}}$ as $k_\perp^2$ goes below $Q^2\sim Q_s^2$ is very slow compared to the decrease of $E_T$ as $k_\perp$ decreases. Thus the dominant transverse energy produced is greater than or equal to $k_\perp \sim Q_s$. However, when $k_\perp > Q_s$ the probability of producing the gluon of $k_\perp$ goes as $Pr(k_\perp)\sim Q_s^2/k_\perp^2$ and those gluons which are produced and have $k_\perp/Q_s>1$ interact too weakly 
with the gluons to contribute to the quark-gluon plasma. The occupation $f(k_\perp)$ being greater than $1/\alpha$ when $k_\perp/Q_s<1$ does not modify our picture of the early stages after a heavy ion collision.

\subsection{Gluon recombination} \label{sec:5.4}

As we have seen earlier, if not for Sudakov effects gluon occupancies could grow arbitrarily large as the gluons momentum fraction of a nucleon in the nucleus becomes small. It has been widely believed that in such extreme limits with so many gluons overlapping each other that gluons would recombine (2 gluons $\to$ 1 gluon) to lower their occupancy. So far we have seen no hint of such an occupancy lowering mechanism in our discussion. Let's probe a bit deeper into this issue in the simple case of a quasiclassical gluon. We have seen
in \eqref{eq:3.5} that the gluon occupancy becomes arbitrarily large as $Q_s^2/k_\perp^2$ becomes large, and we have identified the graphs in Fig.~\ref{fig:4} and Fig.~\ref{fig:5} for the projectile and target frames, respectively, as the dominant contribution to the gluon TMD.

\begin{figure}[htbp]
\centering
\begin{subfigure}[b]{0.3\textwidth}
\centering
\begin{tikzpicture}
\begin{feynman}
\vertex [blob, minimum height=0.5cm, minimum width=0.5cm] (n1) {$1$};
\vertex [left=0.5cm of n1] (nl1);
\vertex [right=0.5cm of n1] (nr1);
\vertex [above=0.2cm of nr1] (ntr1);
\vertex [below=0.2cm of nr1] (nbr1);
\vertex [below right=1.5cm of n1] (b1);
\vertex [right=1.0cm of b1] (b2);
\vertex [right=1.5cm of b2] (b3);
\vertex [below left=1.5cm of b2] (q);

\diagram* {
  (nl1) -- [double] (n1) -- (nr1),
  (ntr1) -- (n1) -- (nbr1),
  (n1) -- [photon, quarter right] (b1) -- [charged boson, edge label=$k$] (b2) -- [charged boson, edge label'=$q+k$] (b3),
  (q) -- [charged scalar, edge label'=$q$] (b2),
};
\end{feynman}
\end{tikzpicture}
\caption{\label{fig:12A}}
\end{subfigure}
\hfill
\begin{subfigure}[b]{0.4\textwidth}
\centering
\begin{tikzpicture}
\begin{feynman}
\vertex [blob, minimum height=0.5cm, minimum width=0.5cm] (n1) {$1$};
\vertex [left=0.5cm of n1] (nl1);
\vertex [right=0.5cm of n1] (nr1);
\vertex [above=0.2cm of nr1] (ntr1);
\vertex [below=0.2cm of nr1] (nbr1);
\vertex [below right=1.5cm of n1] (b1);
\vertex [right=0.125cm of b1] (b2);
\vertex [right=0.25cm of b2] (b3);
\vertex [right=1.7cm of b3] (b4);
\vertex [right=1.5cm of b4] (b5);
\vertex [below left=1.5cm of b4] (q);
\vertex [right=1.3cm of n1, blob, minimum height=0.5cm, minimum width=0.5cm] (n2) {$2$};
\vertex [left=0.5cm of n2] (nl2);
\vertex [right=0.5cm of n2] (nr2);
\vertex [above=0.2cm of nr2] (ntr2);
\vertex [below=0.2cm of nr2] (nbr2);

\diagram* {
  (nl1) -- [double] (n1) -- (nr1),
  (ntr1) -- (n1) -- (nbr1),
  (nl2) -- [double] (n2) -- [double] (nr2),
  (n1) -- [charged boson, quarter right, edge label'=$k-\delta$] (b1) -- [photon] (b3) -- [charged boson, edge label'=$k$] (b4) -- [charged boson, edge label'=$q+k$] (b5),
  (n2) -- [charged boson, edge label'=$l$, in=90, out=255] (b2),
  (b3) -- [charged boson, edge label'=$l-\delta$, out=90, in=-70] (n2),
  (q) -- [charged scalar, edge label'=$q$] (b4),
};
\end{feynman}
\end{tikzpicture}
\caption{\label{fig:12B}}
\end{subfigure}
\hfill
\begin{subfigure}[b]{0.4\textwidth}
\centering
\begin{tikzpicture}
\begin{feynman}
\vertex [blob, minimum height=0.5cm, minimum width=0.5cm] (n1) {$1$};
\vertex [left=0.5cm of n1] (nl1);
\vertex [right=0.5cm of n1] (nr1);
\vertex [above=0.2cm of nr1] (ntr1);
\vertex [below=0.2cm of nr1] (nbr1);
\vertex [below right=1.5cm of n1] (b1);
\vertex [right=0.25cm of b1] (b2);
\vertex [right=1.7cm of b2] (b3);
\vertex [right=1.5cm of b3] (b4);
\vertex [below left=1.5cm of b3] (q);
\vertex [right=1.3cm of n1, blob, minimum height=0.5cm, minimum width=0.5cm] (n2) {$2$};
\vertex [left=0.5cm of n2] (nl2);
\vertex [right=0.5cm of n2] (nr2);
\vertex [above=0.2cm of nr2] (ntr2);
\vertex [below=0.2cm of nr2] (nbr2);
\vertex [below=0.01cm of b2] {$\alpha$};
\vertex [below right=0.25cm of n2] (tmp);
\vertex [below=0.0cm of tmp] {$\beta$};

\diagram* {
  (nl1) -- [double] (n1) -- (nr1),
  (ntr1) -- (n1) -- (nbr1),
  (nl2) -- [double] (n2) -- (nr2),
  (ntr2) -- (n2) -- (nbr2),
  (n1) -- [charged boson, quarter right, edge label'=$k-l$] (b1) -- [photon] (b2) -- [charged boson, edge label'=$k$] (b3) -- [charged boson, edge label'=$q+k$] (b4),
  (n2) -- [charged boson, edge label'=$l$] (b2),
  (q) -- [charged scalar, edge label'=$q$] (b3),
};
\end{feynman}
\end{tikzpicture}
\caption{\label{fig:12C}}
\end{subfigure}
\caption{ \label{fig:12}}
\end{figure}

Let's focus on the case of two nucleons of the nucleus having gluons which interact. In Fig.~\ref{fig:12} we depict the production of a gluon from nucleon 1 as well as the interaction of gluons coming from nucleon 1 and nucleon 2. Let's interpret these graphs in the way we did in Sec.~\ref{sec:3.1} and in Fig.~\ref{fig:5} and then ask the question if recombination could be hidden somewhere here.

In Fig.\ref{fig:12A} one is clearly producing a single soft gluon in the wavefunction of nucleon 1. The virtual contribution of Fig.\ref{fig:12B} takes
probability out of the contribution of Fig.\ref{fig:12A} and this probability is put back in Fig.\ref{fig:12C} but with the gluon being produced at a different momentum than that coming from nucleon 1. The fact that the contributions illustrated in Fig.\ref{fig:12B} and Fig.\ref{fig:12C} have equal and opposite probabilities is just the lack of shadowing that was discussed earlier.

The way we have drawn Fig.\ref{fig:12C} makes it look like a recombination of two gluons into one. However, the straightforward way to evaluate this graph is to focus on the 
\begin{equation} \label{eq:5.5}
    \frac{1}{l^2+i\epsilon}\ \frac{\eta_\beta l_\alpha}{l_- - i\epsilon}
\end{equation}
term in the propagator and to evaluate the graph by distorting the $l_-$-integration into the upper half $l_-$-plane and noting that the $l_-=0$ pole in \eqref{eq:5.5} is 
the only singularity encountered. The $l_\alpha$ then does a simple gauge rotation as suggested by the graph in Fig.~\ref{fig:5}. Thus rather than recombinations we seem to be getting gauge rotations so that the higher interactions in this quasiclassical model simply change the distribution of transverse momenta of the gluons as was already suggested by the lack of shadowing in this model. It may be that one can view a part of Fig.\ref{fig:12C} as a gluon recombination but that part would be intimately tied to other parts of the graph which overall do not look like recombination. Nevertheless it would be very interesting to have a deeper picture of recombination to see if there are actual contributions which take probability out of two gluon states and put that probability into one gluon states thus 
lowering gluon occupancies.

\subsection{Drell-Yan muon production in nucleus-nucleus collisions: coherent and incoherent contributions} \label{sec:5.5}

This final topic does not relate to gluon saturation but rather to quark saturation. It is included here in order to stimulate the discovery of a similar phenomenon involving gluon saturation.

\begin{figure}[htbp]
\centering
\begin{tikzpicture}
\begin{feynman}
\vertex (ntl1);
\vertex [right=0.5cm of ntl1] (ntr1);
\vertex [below=1.5cm of ntl1] (nbl1);
\vertex [below=1.5cm of ntr1] (nbr1);
\vertex [above left=0.5cm of ntl1] (nt1) {$P$};
\vertex [below left=0.5cm of nbl1] (nb1) {$P$};

\vertex [right=1.5cm of ntr1] (tm);
\vertex [right=1.5cm of tm] (ntl2);
\vertex [right=1.5cm of nbr1] (bm);
\vertex [right=1.5cm of bm] (nbl2);

\vertex [above=0.5cm of bm, blob, minimum height=0.5cm, minimum width=0.5cm] (blob) {};

\vertex [right=0.5cm of ntl2] (ntr2);
\vertex [right=0.5cm of nbl2] (nbr2);
\vertex [above right=0.5cm of ntr2] (nt2) {$P'$};
\vertex [below right=0.5cm of nbr2] (nb2) {$P'$};

\vertex [below=0.75cm of ntl1] (ml);
\vertex [left=0.5cm of ml] (ll);
\vertex [below=0.75cm of ntr2] (mr);
\vertex [right=0.5cm of mr] (rr);

\diagram* {
  (ntl1) -- [half left] (ntr1) -- (nbr1) -- [half left] (nbl1) -- (ntl1),
  (nt1) -- [double] (ntl1),
  (nb1) -- [double] (nbl1),
  (ntl2) -- [half left] (ntr2) -- (nbr2) -- [half left] (nbl2) -- (ntl2),
  (nt2) -- [double] (ntr2),
  (nb2) -- [double] (nbr2),
  (ntr1) -- [anti fermion, edge label=$p$] (tm) -- [fermion, edge label=$q-p$] (ntl2),
  (nbr1) -- [fermion, edge label'=$p$] (bm) -- [anti fermion, edge label'=$q-p$] (nbl2),
  (bm) -- [photon] (blob) -- [charged boson, edge label=$q$] (tm),
};

\diagram* {
  (ll) -- [very thick] (rr),
};
\end{feynman}
\end{tikzpicture}
\caption{ \label{fig:13}}
\end{figure}

Drell-Yan muon pair production in nucleus-nucleus collisions, with the transverse momentum of the $\mu$-pair fixed at $q_\perp$ and the mass at $Q$ with $Q^2=q_\mu q_\mu$ is illustrated in Fig.~\ref{fig:13}. The momentum $q_\mu$ is also the momentum of the $\gamma^*$, produced by the quark-antiquark annihilation, which then decays into the $\mu$-pair. The left and right-hand 
parts of Fig.~\ref{fig:13} are then the quark and antiquark TMD's for the nuclei which are then integrated over $p_\perp$. When $q_\perp < Q_s$ but $Q>Q_s$ an interesting phenomenon occurs. In Fig.~\ref{fig:13} the inelastic TMD's are shown but when $q_\perp<Q_s$ the coherent TMD's also come in as illustrated in Fig.~\ref{fig:14} as well as the product of an inelastic TMD for the quark (antiquark) times a coherent TMD for the antiquark (quark). In the coherent TMD's in Fig.~\ref{fig:14} the nuclei will be broken up in the final state but as far as the $\mu$-pair cross section is concerned that breakup is unimportant. Well in the saturation region we expect equal contributions from: inelastic TMD $\cdot$ inelastic TMD (Fig.~\ref{fig:13}), elastic TMD $\cdot$ elastic TMD (Fig.~\ref{fig:14}), elastic TMD $\cdot$ inelastic TMD and 
inelastic TMD $\cdot$ elastic TMD making the total $\mu$-pair production 4 times the expectation from the usual Drell-Yan calculation.

\begin{figure}[htbp]
\centering
\begin{tikzpicture}
\begin{feynman}
\vertex [blob, minimum height=0.5cm, minimum width=0.5cm] (n1) {};
\vertex [above=0.5cm of n1] (nt1);
\vertex [below=0.5cm of n1] (nb1);
\vertex [below=1.5cm of n1, blob, minimum height=0.5cm, minimum width=0.5cm] (n2) {};
\vertex [above=0.5cm of n2] (nt2);
\vertex [below=0.8cm of n2] (nb2) {$P$};
\vertex [right=1.5cm of n1] (tl);
\vertex [right=1.5cm of tl] (tm);
\vertex [right=1.5cm of tm] (tr);
\vertex [right=1.5cm of tr, blob, minimum height=0.5cm, minimum width=0.5cm] (n3) {};
\vertex [above=0.5cm of n3] (nt3);
\vertex [below=0.5cm of n3] (nb3);
\vertex [right=1.5cm of n2] (bl);
\vertex [right=1.5cm of bl] (bm);
\vertex [right=1.5cm of bm] (br);
\vertex [right=1.5cm of br, blob, minimum height=0.5cm, minimum width=0.5cm] (n4) {};
\vertex [above=0.5cm of n4] (nt4);
\vertex [below=0.8cm of n4] (nb4) {$P'$};

\vertex [above=0.5cm of bm, blob, minimum height=0.5cm, minimum width=0.5cm] (blob) {};

\vertex [below=0.75cm of n1] (ml);
\vertex [left=0.5cm of ml] (ll);
\vertex [below=0.75cm of n3] (mr);
\vertex [right=0.5cm of mr] (rr);

\diagram* {
  (n1) -- [gluon] (tl) -- [anti fermion, edge label=$p$] (tm) -- [fermion, edge label=$q-p$] (tr) -- [gluon] (n3),
  (n2) -- [gluon] (bl) -- [fermion, edge label'=$p$] (bm) -- [anti fermion, edge label'=$q-p$] (br) -- [gluon] (n4),
  (bm) -- [charged boson, edge label=$q$] (blob) -- [photon] (tm),
  (tl) -- (bl),
  (tr) -- (br),
  (nt1) -- [double] (n1) -- [double] (nb1),
  (nt2) -- [double] (n2) -- [double] (nb2),
  (nt3) -- [double] (n3) -- [double] (nb3),
  (nt4) -- [double] (n4) -- [double] (nb4),
};

\diagram* {
  (ll) -- [very thick] (rr),
};
\end{feynman}
\end{tikzpicture}
\caption{ \label{fig:14}}
\end{figure}

In the case of gluon TMD's, Higgs production can in principal come from coherent gluon TMD's as shown in Fig.~\ref{fig:15} for a Higgs of transverse momentum $q$ and with $q_\perp/Q_s\ll 1$. However, I doubt that this example is realistic because the large mass of the Higgs will make Sudakov effects extremely large. In any case it is an interesting challenge to find some reasonably realistic case where inelastic and coherent gluon TMD's both contribute comparably.

\begin{figure}[htbp]\centering
\begin{tikzpicture}
\begin{feynman}
\vertex [blob, minimum height=0.5cm, minimum width=0.5cm] (n1) {};
\vertex [above=0.5cm of n1] (nt1);
\vertex [below=0.5cm of n1] (nb1);
\vertex [below=1.5cm of n1, blob, minimum height=0.5cm, minimum width=0.5cm] (n2) {};
\vertex [above=0.5cm of n2] (nt2);
\vertex [below=0.8cm of n2] (nb2) {$P$};
\vertex [right=1.5cm of n1] (tl);
\vertex [right=1.5cm of tl] (tm);
\vertex [right=1.5cm of tm] (tr);
\vertex [right=1.5cm of tr, blob, minimum height=0.5cm, minimum width=0.5cm] (n3) {};
\vertex [above=0.5cm of n3] (nt3);
\vertex [below=0.5cm of n3] (nb3);
\vertex [right=1.5cm of n2] (bl);
\vertex [right=1.5cm of bl] (bm);
\vertex [right=1.5cm of bm] (br);
\vertex [right=1.5cm of br, blob, minimum height=0.5cm, minimum width=0.5cm] (n4) {};
\vertex [above=0.5cm of n4] (nt4);
\vertex [below=0.8cm of n4] (nb4) {$P'$};

\vertex [above=0.75cm of bm] (mm);

\vertex [below=0.75cm of n1] (ml);
\vertex [left=0.5cm of ml] (ll);
\vertex [below=0.75cm of n3] (mr);
\vertex [right=0.5cm of mr] (rr);

\diagram* {
  (n1) -- [gluon] (tl) -- [photon] (tm) -- [photon] (tr) -- [gluon] (n3),
  (n2) -- [gluon] (bl) -- [photon] (bm) -- [photon] (br) -- [gluon] (n4),
  (bm) -- [fermion, edge label'=$q$] (mm) -- (tm),
  (tl) -- [photon] (bl),
  (tr) -- [photon] (br),
  (nt1) -- [double] (n1) -- [double] (nb1),
  (nt2) -- [double] (n2) -- [double] (nb2),
  (nt3) -- [double] (n3) -- [double] (nb3),
  (nt4) -- [double] (n4) -- [double] (nb4),
};

\diagram* {
  (ll) -- [very thick] (rr),
};
\end{feynman}
\end{tikzpicture}
\caption{ \label{fig:15}}
\end{figure}

\section*{Acknowledgements}

I have worked on and off with saturation over many years and have had useful and inspiring converstaions with many colleagues including J.-P. Blaizot, L.~ Frankfurt and M.~ Strikman, E.~ Iancu, Y.~ Kovchegov, E.~ Levin, L.~ McLerran, S.~ Munier, D.~ Triantofyllopoulos, R.~ Venugopalan, B.~ Xiao and F.~ Yuan. THis work was supported in part through a DOE grant DE-SC0011941.

\bibliographystyle{unsrt}
\bibliography{refs}

\end{document}